\documentclass[a4paper]{article}
\usepackage{framed}

\title{ Improved Approximation Algorithms for (1,2)-TSP and Max-TSP Using Path Covers in the Semi-Streaming Model}

\author{Sharareh Alipour 
	\\Tehran Institute for Advanced Studies (TeIAS), Khatam University 
	\\\texttt{sharareh.alipour@gmail.com}  
	\and
	Ermiya Farokhnejad
	\\University of Warwick
	\\\texttt{ermiya.farokhnejad@warwick.ac.uk}
	\and
	Tobias M\"{o}mke\thanks{Partially supported by DFG Grant 439522729 (Heisenberg-Grant)}
	\\University of Augsburg
	\\\texttt{moemke@informatik.uni-augsburg.de} }

\date{}

\usepackage{graphicx} 
\usepackage{fancyhdr}
\usepackage{blindtext}
\usepackage{amsmath, amsthm}
\usepackage{amssymb}
\usepackage{mathtools}
\usepackage{pdfpages}
\usepackage{anyfontsize}
\usepackage{xstring}
\usepackage{xargs}
\usepackage{makeidx}
\usepackage{cancel}
\usepackage{chngcntr}
\usepackage{caption}
\usepackage{subcaption}
\usepackage{float}
\usepackage{pgf,tikz,pgfplots}
\usepackage{latexsym}
\usepackage{verbatim}
\usepackage{amsmath,array}
\usepackage{tkz-graph}
\usepackage{tkz-euclide}
\usepackage{bookmark}
\usepackage[toc]{appendix}
\usetikzlibrary{positioning,chains,fit,shapes,calc,decorations.markings,decorations.pathreplacing,arrows.meta,snakes}

\usepackage[utf8]{inputenc}

\usepackage{amsthm}
\usepackage{amsmath}
\usepackage{amssymb}
\usepackage{enumerate}
\usepackage{graphicx}
\usepackage[margin=1in]{geometry}
\usepackage{dsfont}
\usepackage{booktabs} 
\usepackage{tabularx}
\usepackage{multirow}
\usepackage{enumitem}
\usepackage{tablefootnote}

\usepackage{listings}
\usepackage{hyperref}
\usepackage{cleveref}

\usepackage{subcaption}

\hypersetup{
	colorlinks=true,
	linkcolor=blue,
	filecolor=magenta,      
	urlcolor=cyan,
}
\usepackage{algorithmicx,algorithm}
\usepackage{algpseudocode}

\usepackage{bidicontour}

\newcounter{mytheorem}

\makeatletter

\renewcommand\themytheorem{\thesection.\arabic{mytheorem}}

\newenvironment{mytheorem}[1]{\refstepcounter{mytheorem} 
	\textbf{Theorem \themytheorem \ #1}\itshape
}{}

\newenvironment{mylemma}[1]{\refstepcounter{mytheorem} 
	\textbf{Lemma \themytheorem \ #1}\itshape
}{}

\newenvironment{mycorr}[1]{\refstepcounter{mytheorem}
	\textbf{Corollary \themytheorem \ #1}\itshape
}{}

\newenvironment{myproof}{\noindent\textit{Proof.}}{\hfill $\square$}

\hypersetup{ colorlinks, citecolor=blue}

\usepackage{xspace}
\newcommand{\tsp}{\mbox{$(1,2)$-TSP}\xspace}

\begin{document}

\maketitle

\begin{abstract}

We investigate semi-streaming algorithms for the Traveling Salesman Problem (TSP). Specifically, we focus on a variant known as the
\tsp, where the distances between any two vertices are either \emph{one} or \emph{two}. Our primary emphasis is on the closely related Maximum Path Cover Problem, which aims to find a collection of vertex-disjoint paths that covers the maximum number of edges in a graph. We propose an algorithm that, for any $\epsilon > 0$, achieves a
$(\frac{2}{3}-\epsilon)$-approximation of the maximum path cover size for an
$n$-vertex graph, using
$\text{poly}(\frac{1}{\epsilon})$ passes. This result improves upon the previous
$\frac{1}{2}$-approximation by Behnezhad et al.~\cite{beh} in the semi-streaming model. Building on this result, we design a semi-streaming algorithm that constructs a tour for an instance of \tsp with an approximation factor of $(\frac{4}{3} + \epsilon)$, improving upon the previous $\frac{3}{2}$-approximation factor algorithm by Behnezhad et al.~\cite{beh}\footnote{Although Behnezhad et al. do not explicitly state that their algorithm works in the semi-streaming model, it is easy to verify.}.

Furthermore, we extend our approach to develop an approximation algorithm for the Maximum TSP (Max-TSP), where the goal is to find a Hamiltonian cycle with the maximum possible weight in a given weighted graph
$G$. Our algorithm provides a
$(\frac{7}{12} - \epsilon)$-approximation for Max-TSP in
$\text{poly}(\frac{1}{\epsilon})$ passes, improving on the previously known
$(\frac{1}{2}-\epsilon)$-approximation obtained via maximum weight matching in the semi-streaming model.
\end{abstract}

\newpage

 	\section{Introduction}

The Traveling Salesman Problem (TSP) is a fundamental problem in combinatorial optimization. Given a graph $G = (V, E)$ with distances assigned to the edges, the objective is to find a Hamiltonian cycle with the lowest possible cost. The general form of TSP is known to be inapproximable unless $\mathsf{P} = \mathsf{NP}$~\cite{gonz}. Consequently, research often focuses on specific types of distance functions, particularly the metric TSP, where distances satisfy the triangle inequality. Two notable metric versions of TSP are the graphic TSP, where distances correspond to the shortest path lengths in an unweighted graph, and \tsp, a variant of TSP with distances restricted to either \emph{one} or \emph{two}~\cite{anna, chen1, chen, czumaj, shayan1, anna2, Book1975-BOORMK-2, mnich, tobias, mucha, seb, zhongapproximation2, zhong2021approximation}.

Most research is conducted within the classic centralized model of computation. However, with the surge of large data sets in various real-world applications (as reviewed in \cite{DrakeH03}), there is a growing demand for algorithms capable of handling massive inputs. For very large graphs, classical algorithms are not only too slow but also suffer from excessive space complexity. When a graph's size exceeds the memory capacity of a single machine, algorithms that rely on random access to the data become impractical, necessitating alternative computational models.
One such model that has gained significant attention recently is the graph stream model, introduced by Feigenbaum et al.~\cite{FeigenbaumKMSZ052, FeigenbaumKMSZ05}. In this model, edges of the graph are not stored in memory but arrive sequentially in a stream, requiring processing in that order. The challenge is to design algorithms that use minimal space and ideally make only a small constant number of passes over the stream. A widely studied variant of this is the semi-streaming model. In the semi-streaming model, as outlined by Feigenbaum et al.~\cite{FeigenbaumKMSZ05}, we consider a graph $G$ with $n$ vertices. The algorithm processes the graph's edges as they arrive in the stream and aims to compute results with minimal passes while using limited memory, constrained to $\tilde{O}(n) := O(n \cdot \text{polylog}(n))$.

It is straightforward to design a deterministic one-pass streaming algorithm to compute the cost of a Minimum Spanning Tree (MST) exactly, even in graph streams, which in turn immediately provides an $\tilde{O}(n)$ space algorithm to estimate TSP cost within a factor of 2. Thus, in the semi-streaming regime, the key challenge is to estimate TSP cost within a factor that is strictly better than 2. Recently, Chen, Khanna, and Tan \cite{chenete} proposed a deterministic two-pass 1.96-approximation factor algorithm for metric TSP cost estimation in the semi-streaming model. For the case of \tsp, using the approach of Behnezhad et al.~\cite{beh}, it is possible to provide a $1.5$-approximation factor algorithm in the semi-streaming model. In \cite{beh}, authors presented a sub-linear version of their algorithm; however, it is straightforward to implement their algorithm in the semi-streaming model.

In section 11 of \cite{beh}, the authors showed a reduction from (1,2)-TSP to maximum matching and stated that achieving better approximation than $1.5$ for (1,2)-TSP in the sub-linear model, solves an important open problem in sub-linear maximum matching.
Considering the same reduction in semi-streaming model shows achieving non-trivial approximations for (1,2)-TSP in semi-streaming model is challenging.
Since the maximum matching problem is studied in the semi-streaming model extensively, and there are important open problems there, the following question naturally arises.
\begin{framed}
	\textbf{Question.}
	What is the trade off between the approximation ratio and the number of passes for (1,2)-TSP in the semi-streaming model?
\end{framed}

\textbf{Maximum Path Cover}:
In an unweighted graph $G$, a subset of edges is called a \textit{path cover} if it forms a union of vertex-disjoint paths. A maximum path cover (MPC\footnote{Throughout this paper we use this acronym for 'Maximum Path Cover'. Please note that we do \textbf{not} refer to the common abbreviation for 'Massively Parallel Computation'.}) in an unweighted graph is a path cover with the maximum number of edges (not paths) among all possible path covers in the graph. The problem of finding an MPC is known to be $\mathsf{NP}$-complete. It is straightforward to see that a maximum matching provides a $1/2$-approximation for MPC. Therefore, computing a maximal matching, which is a $1/2$-approximation for maximum matching, yields a $1/4$-approximate solution for MPC.

Behnezhad et al.~\cite{beh} developed a $1/2$-approximate MPC algorithm, which provides a $1.5$-approximate solution for \tsp. Their algorithm can be implemented in one pass within the semi-streaming model using $\tilde{O}(n)$ space to return the cost, and in two passes if the approximate solution itself is required. Our primary contribution is an improvement in the approximation factor of their algorithm.

    \begin{framed}
             \textbf{Result 1 
 (Formally as \hyperref[main1]{Theorem \ref*{main1}}).}
            For a given unweighted graph $G$, there is a semi-streaming algorithm that returns a $(\frac{2}{3} - \epsilon)$-approximation of MPC in $\text{poly}(\frac{1}{\epsilon})$ passes.
    \end{framed}

\textbf{\tsp}:
The classical problem \tsp is well-studied and known to be $\mathsf{NP}$-hard \cite{Book1975-BOORMK-2}, and even $\mathsf{APX}$-hard \cite{PAPADIMITRIOU1991425}.
One can easily observe that in an instance of \tsp, the optimal tour is almost the same as finding the MPC of the induced subgraph on edges with weight 1 and then joining their endpoints with edges with weight 2, except for a possible difference of 1 (in the case that there exists a Hamiltonian cycle all of whose edges have weight 1). A simple computation shows that if one can find a set of vertex-disjoint paths that is at least $\alpha$ times the optimal size ($\alpha \leq 1$), then one can also find a tour whose cost is no more than $(2 - \alpha)$ times the optimal cost for \tsp.
Thus, Result 1 implies the following result.
    \begin{framed}
        \textbf{Result 2 (Formally as \hyperref[main2]{Theorem \ref*{main2}}).}
            For an instance of \tsp, there is a semi-streaming algorithm that returns a ($\frac{4}{3}+\epsilon$)-approximation of \tsp in $\text{poly}(\frac{1}{\epsilon})$ passes.
    \end{framed}

In the second part of the paper, we examine  Max-TSP in the semi-streaming model.

\textbf{Max-TSP}:
For a given complete weighted graph $G$, the goal of Max-TSP is to find a Hamiltonian cycle such that the sum of the weights of the edges in this cycle is maximized.
 
It is evident that a maximum weighted matching provides a $\frac{1}{2}$-approximation for the cost of Max-TSP. Consequently, the result of Huang and Saranurak \cite{huangS23}, which computes a $(1-\epsilon)$-approximate maximum weight matching in the semi-streaming model, yields a ($\frac{1}{2}-\epsilon$)-approximation for Max-TSP. In this paper, we improve this bound to $\frac{7}{12} - \epsilon$. Our result is as follows.

  \begin{framed}
            
        \textbf{Result 3 (Formally as \hyperref[main3]{Theorem \ref*{main3}}).}
            For a given weighted graph $G$, there is a semi-streaming algorithm that returns a $(\frac{7}{12} - \epsilon)$-approximation of Max-TSP in $\text{poly}(\frac{1}{\epsilon})$ passes.
    \end{framed}
To the best of our knowledge, this is the first non-trivial approximation algorithm for Max-TSP in the semi-streaming model. 
\subsection*{Further related work}
Our approach for computing MPC, \tsp and Max-TSP mainly uses the subroutines for computing maximum matching in unweighted graphs and maximum weight matching in weighted graphs.

In the semi-streaming model,
Fischer, Mitrovic and Uitto \cite{FischerMU22} gave a $(1-\epsilon)$-approximation for the maximum matching problem in $\text{poly}(1/\epsilon)$ passes. This result was an improvement over the $(1/\epsilon)^{O(1/\epsilon)}$ passes algorithm by McGregor \cite{McGregor05}.

For the maximum weight matching in the semi-streaming model, Paz and  Schwartzman gave a simple deterministic single-pass $(1/2-\epsilon)$-approximation algorithm \cite{paz20182+}.
Gamlath, Kale, Mitrovic, and Svensson gave 
a $(1-\epsilon)$-approximation streaming algorithm that uses $O_{\epsilon}(1)$ passes and
$O_{\epsilon}(n \cdot \text{poly}(\log n))$
memory. This was the first $(1-\epsilon)$-approximation streaming algorithm for weighted matching
that uses a constant number of passes (only depending on $\epsilon$) \cite{GamlathKMS19}.
Also, Huang and Su in \cite{huangS23}, gave a deterministic $(1-\epsilon)$-approximation for maximum weighted matching using $\text{poly}(1/\epsilon)$ passes in the semi-streaming model.
When $\epsilon$ is smaller than a constant $O(1)$ but at least $1/ \log^{o(1)} n$, their algorithm is more efficient than \cite{GamlathKMS19}.
	
\subsection{Notation}\label{not-pre}
Let $G$ be a simple graph. We denote the set of vertices and edges of $G$ by $V(G)$ and $E(G)$ respectively.
We also denote the maximum matching size in $G$ by $\mu(G)$ and the size of the MPC in $G$ by $\rho(G)$.

For a subset of edges $T \subseteq E(G)$, we denote $G/T$ as the contraction of $G$ on $T$, which is the graph derived by repeatedly removing edges of $T$ (it is well-known that the order does not matter) from the graph and merging its endpoints to be a single node in the new graph. Note that after contraction the graph might have parallel edges, but this does not interfere with our algorithm.
In the weighted case, if $w(e)$ is the weight of edge $e$, then we define $w(T)$ to be the sum of weights of the elements of $T$ i.e. $\sum_{e \in T} w(e)$.
Let $P=(u_1,u_2,\ldots, u_k)$ be a path of length $k-1$ ($u_i \in V$ for $1 \leq i \leq k$). We call $u_1$ and $u_k$ \textit{end points} of $P$ and $u_i$ for $2 \leq i \leq k-1$ \textit{middle points} of $P$.

    \section{Technical Overview and Our Contribution}\label{sec:technique}

\newcommand{\MM}{\text{MM}_\epsilon}

\newcommand{\MWM}{\text{MWM}_\epsilon}

We propose a simple algorithm that constructs a path cover with an approximation factor of almost $\frac{2}{3}$ for MPC. This new algorithm merely depends on basic
operations and computing matching and approximate matching.

Our algorithm for MPC is as follows. Assume that $\text{MM}{\epsilon}$ is a $(1-\epsilon)$-approximation algorithm for computing maximum matching in an unweighted graph $G$. We use $\text{MM}{\epsilon}$ as a subroutine in our algorithm, which has two phases.
In the first phase, we run $\text{MM}{\epsilon}$ to compute a $(1-\epsilon)$-approximate maximum matching, denoted by $M_1$. In the next phase, we contract all edges in $M_1$ to obtain a new graph $G' = G/M_1$. Then, we compute a $(1-\epsilon)$-approximate maximum matching, denoted by $M_2$, for $G'$ using $\text{MM}{\epsilon}$. Finally, we return the edges of $M_1$ and $M_2$ as a path cover for $G$ (see \hyperref[alg1]{Algorithm 1}).

We will show that the output of our algorithm is a collection of vertex-disjoint paths, i.e.,~a valid path cover (see \hyperref[correct-alg1]{Lemma \ref*{correct-alg1}}).

The algorithm is simple, but proving that its approximation factor is $\frac{2}{3} - \epsilon$ is challenging. As a warm-up, it is straightforward to see that by computing a maximum matching in the first phase, we achieve a $\frac{1}{2}$-approximate MPC. However, the challenge lies in the second phase, which helps to improve the approximation factor.

Now we explain the idea of our proof to find the approximation factor of \hyperref[alg1]{Algorithm 1}. For the matching $M_1$ in graph $G$, we provide a lower bound for $\mu(G/M_1)$.
We show that if we consider a maximum path cover $P^*$ and contract $P^*$ on $M_1$, the contracted graph becomes a particular graph in which we can find a lower bound on the size of its maximum matching.
Let $M_2$ be a maximum matching of $G/M_1$, then this results in a lower bound for $|M_2|$. 
Finally, we exploit this lower bound for $|M_2|$ together with a lower bound for $|M_1|$, to come up with the approximation factor of \hyperref[alg1]{Algorithm 1}. 

We explain how to implement this algorithm in the semi-streaming model, achieving an improved approximation factor for \tsp within this model.

For the Max-TSP, we use a similar algorithm, except we compute maximum weight matching instead of maximum matching (see \hyperref[alg3]{Algorithm 3}). By computing the approximation factor of this algorithm, we provide a non-trivial approximation algorithm for Max-TSP in the semi-streaming model. Despite the extensive study of the weighted version of the maximum matching problem, Max-TSP has not been studied extensively in the literature within the semi-streaming model. One reason could be that it is not possible to extend the approaches for the unweighted version to the weighted version. Fortunately, we can extend our algorithm to the weighted version and improve the approximation factor of Max-TSP in the semi-streaming model. However, our method for analyzing the approximation factor of \hyperref[alg1]{Algorithm 1} does not apply to the weighted version, so we present a different proof approach for computing the approximation factor of \hyperref[alg3]{Algorithm 3}.

    \section{Improved Approximation Factor Semi-Streaming Algorithm for MPC}\label{sec:alg1}

In \hyperref[alg1]{Algorithm 1}, we presented our novel algorithm for MPC. This section provides an analysis of its approximation factor, followed by a detailed explanation of its streaming implementation.

\begin{algorithm}
	\caption{Approximating maximum path cover on a graph $ G $.}
	\begin{algorithmic}[1]\label{alg1}
		\State Run $ \text{MM}_{\epsilon} $ on $ G $ to find a matching  $M_1$.
		\State Contract $ G $ on $ M_1 $ to get a new graph $ G^{\prime} = G/M_1 $.
		\State Run $ \text{MM}_{\epsilon} $ on $ G^{\prime} $ to find another matching $ M_2 $.
		\State \Return $ M_1 \cup M_2 $.
	\end{algorithmic}	
\end{algorithm}

We start by proving the correctness of this algorithm.

\begin{mylemma}{}\label{correct-alg1}
	If $ M_1 $ and $M_2$ are the matchings obtained in \hyperref[alg1]{Algorithm 1}, then $M_1 \cup M_2$ forms a path cover for $G$.
\end{mylemma}

\begin{myproof}
	We claim that $M_1 \cup M_2$ is a vertex-disjoint union of paths of length $1,2$ or $3$. As a result, it is a path cover. Suppose $ M_1 = \{ u_1v_1, u_2v_2, \ldots , u_kv_k \}$. Let us denote the vertices of $ G/ M_1$ by
	$ \{ (uv)_1, (uv)_2, \ldots , (uv)_k, w_1,w_2,\ldots, w_l \} $ where $(uv)_i$ represents the vertices $u_i$ and $v_i$, merged in the contracted graph $G/M_1$ and $w_j$'s for $1 \leq j \leq l$ are the rest of the vertices. Let $xy \in M_2$ be an arbitrary edge. By symmetry between $x$ and $y$, there are three cases as follows:
	\begin{enumerate}
		\item $x,y \in \{w_1, w_2,\ldots, w_l\}$.\\
		In this case, $ x $ and $ y $ are intact vertices after contraction, which means there are no edges in $ M_1 $ adjacent to $x $ and $ y$. Since $xy \in M_2$ and $M_2$ is a matching, there are no other edges in $ M_1 \cup M_2 $ adjacent to $ x $ and $ y $ in $ G $. As a result, $xy$ would be a path of length $1$ in $M_1 \cup M_2$.
		\item $ x \in \{ (uv)_1, (uv)_2, \ldots , (uv)_k \} $ and $ y \in \{w_1, w_2,\ldots, w_l\}$.\\
		In this case, $ x= (uv)_i $ for some $ 1 \leq i \leq k$. As a result, $ xy $ would be $ u_iy $ or $ v_iy $ in $ G $. By symmetry, assume that $xy=u_iy$ in $G$. Since $M_1$ is a matching, no other edges in $M_1$ are adjacent to $u_i$ and $v_i$. No edge in $M_1$ is adjacent to $y$. Since $M_2$ is a matching, the only edge in $M_2$ adjacent to at least one of $u_i,v_i$ and $y$ in $G$ is $xy=u_iy$. Finally, we can see that $(v_i,u_i,y)$ is a path of length $2$ in $M_1 \cup M_2$.
		\item $x,y \in \{ (uv)_1, (uv)_2, \ldots , (uv)_k \}$.\\
		In this case, let $ x=(uv)_i$ and $y=(uv)_j$ and by symmetry, assume that $xy$ is the edge connecting $u_i$ to $u_j$ in $ G $. Since $M_1$ is a matching, the only edges in $M_1$ adjacent to at least one of $u_i,u_j,v_i,v_j$ are $u_iv_i$ and $u_jv_j$. Since $M_2$ is a matching, the only edge in $M_2$ adjacent to at least one of $u_i,u_j,v_i,v_j$ is $(uv)_i(uv)_j$. As a result, $(v_i,u_i,u_j,v_j)$ would be a path of length $3$ in $M_1 \cup M_2$.
	\end{enumerate}
	So, $M_1 \cup M_2$ is a union of vertex-disjoint paths of length $1,2$ or $3$.
\end{myproof}

    \subsection{Analysis of the Approximation Factor of \hyperref[alg1]{Algorithm 1}}\label{sec:apprx}

We start with a simple and basic lemmas that is crucial in our proof.

\begin{mylemma}{}\label{mu-rho}
    Let $ G $ be an arbitrary graph.
    We have:
    $$  \rho(G) \geq \mu(G) \geq  \frac{1}{2} \rho(G).$$
\end{mylemma}

\begin{myproof}
    Since every matching is a path cover, we have  $\rho(G) \geq \mu(G)$. 
    Also, given an MPC, we can select every other edge in this MPC to obtain a matching that contains at least half of its edges, which implies $  \mu(G) \geq \frac{1}{2} \rho(G) $.
\end{myproof}

\begin{mycorr}{}\label{approx-M1}
    If $M$ is a $(1-\epsilon)$-approximation of maximum matching in graph $G$, then 
    $$
      |M| \geq \frac{1}{2}(1-\epsilon) \rho(G) .  
    $$
\end{mycorr}

We now present a lemma regarding the size of the maximum matching in a specific type of graph. This lemma may be of independent interest. We include its proof in \hyperref[appendix1]{Section \ref*{appendix1}}.
In this paper we utilize this lemma on $G/M_1$ to derive a lower bound for $|M_2|$.
 
\begin{mylemma}{}\label{E-deg4}
    Assume $G$ is a graph without loops such that each vertex $v$ of $G$ has degree $1$,$2$, or $4$. If $V_4(G)$ denotes the set of vertices of degree $4$ in $G$, then we have
    $$
        \mu(G) \geq \frac{|E(G)| - |V_4(G)|}{3}.
    $$
\end{mylemma}

Using above lemma, we have the main lemma of this section as follows.

\begin{mylemma}{}\label{mu>(rho-M)/3}
    If $M$ is an arbitrary matching in a graph $G$, then
    $$  \mu(G/M) \geq \frac{\rho(G) - |M|}{3}  .$$
\end{mylemma}

\begin{myproof}
Assume $P^*$ is a maximum path cover in $G$ such that $P^* \cap M$ is maximal.
We claim that every $e \in M \setminus P^*$ connects two middle points of $P^*$. 
The proof of this claim follows from a case by case argument.
For the sake of contradiction, assume $ e = uv  \in M \setminus P^* $ does not connect two middle points of $P^*$. We have three cases for $u$ and $v$ as follows.
    
 \begin{itemize}
     \item None of $u$ and $v$ belong to $P^*$ (case 1).
     \item Exactly one of them (say $u$) belongs to $P^*$. Then, $u$ is an end point (case 2), or $u$ is a middle point (case 3).
     \item Both $u$ and $v$ belong to $P^*$. Then, we have two sub cases.
     \subitem
     $u$ and $v$ are on different paths. Then either they are both end points (case 4), or one is a middle point (say $u$) and the other one is an end point (case 5). Note that we have considered that both of $u$ and $v$ are not middle points at the same time.
     \subitem
     $u$ and $v$ belong to the same path. Then, either they are both end points (case 6), or one is a middle point (say $u$) and the other one is an end point (case 7). Note that we have assumed both of them are not middle points at the same time.
 \end{itemize}

Now, we explain each case in detail. 
\begin{enumerate}
    \item Neither $u$ nor $v$ belongs to  $\tilde{P}$.\\
    This case is impossible because $P^* + e$ is a path cover with a size larger than $|P^*|$, which is in contradiction with  $P^*$ being MPC (see \hyperref[fig-approx-1/2+1/2s-1]{Figure \ref*{fig-approx-1/2+1/2s-1}}).

    \item $ u $ is an end point of a path in $ P^* $ and $ v $ is not contained in $ P^* $.\\
    Again, this case is impossible since $ P^*+e$ is a path cover with a size larger than $|P^*|$, which is in contradiction with $P^*$ being MPC (see  \hyperref[fig-approx-1/2+1/2s-2]{Figure \ref*{fig-approx-1/2+1/2s-2}}).
      
    \item $ u $ is a middle point of a path in $ P^* $ and $ v $ is not contained in $ P^* $.\\
    Let $ (p_1,p_2,\ldots, p_k) $ be the path in $ P^* $ containing $ u = p_ i $.
    Replace $ P^* $ by $ P^* - p_{i-1}p_i+e$ which is an MPC of $G$ (see  \hyperref[fig-approx-1/2+1/2s-3]{Figure \ref*{fig-approx-1/2+1/2s-3}}).
    Since $ e \in M $, we have $ p_{i-1}p_i \notin M $. Therefore, $ |\tilde{P} \cap M| $ increments. This is in contradiction with $|P^* \cap M|$ being maximal.

    \item $ u $ and $ v $ are end points of different paths in $ P^* $.\\
    In this case, let $ (p_1,p_2,\ldots, p_k) $ and $ (q_1,q_2,\ldots, q_l) $ be the paths in $ P^* $ containing $ u = p_ 1 $ and $ v=q_1 $, respectively.
    $ P^*+e$ would be a path cover of size greater than $|P^*|$ which is in contradiction with $P^*$ being MPC (see  \hyperref[fig-approx-1/2+1/2s-4]{Figure \ref*{fig-approx-1/2+1/2s-4}}).

    \item $u$ and $v$ are the middle and end points of different paths in $P^*$, respectively.\\
    In this case, let $ (p_1,p_2,\ldots, p_k) $ and $ (q_1,q_2,\ldots, q_l) $ be the paths in $ P^* $ containing $ u = p_ i $ and $ v = q_1 $ respectively.
    Replace $ P^* $ by $ P^* - p_{i-1}p_i +e$ which is an MPC of $G$ (see  \hyperref[fig-approx-1/2+1/2s-6]{Figure \ref*{fig-approx-1/2+1/2s-6}}). Since $ e \in M $, we have $ p_{i-1}p_i \notin M $. Therefore, $ |P^* \cap M| $ increments. This is in contradiction with $|P^* \cap M|$ being maximal.
      
    \item $ u $ and $ v $ are end points of the same path in $ P^* $.\\
    In this case, let $ (p_1,p_2,\ldots, p_k) $ be the path in $ P^* $ containing $ u = p_ 1 $ and $ v=p_k $.
    Replace $ P^*  $ by $ P^* - p_1p_2 +e$ which is an MPC of $G$ (see  \hyperref[fig-approx-1/2+1/2s-7]{Figure \ref*{fig-approx-1/2+1/2s-7}}). Since $ e \in M $ we have $ p_{1}p_2 \notin M $. Therefore, $ |P^* \cap M| $ increments. This is in contradiction with $|P^* \cap M|$ being maximal.

    \item $u$ and $v$ are the middle and end points of the same path in $P^*$, respectively.\\
    In this case, let $ (p_1,p_2,\ldots, p_k) $ be the path in $ P^* $ containing $ u = p_ i $ and $ v = p_1 $ (since $ e \notin P^* $, we have $ 2 < i $).
    Replace $ P^* $ by $ P^* - p_{i-1}p_i +e$ which is an MPC of $G$ (see  \hyperref[fig-approx-1/2+1/2s-9]{Figure \ref*{fig-approx-1/2+1/2s-9}}). Since $ e \in M $, we have $ p_{i-1}p_i \notin M $. Therefore, $ |P^* \cap M| $ increments. This is in contradiction with $|P^* \cap M|$ being maximal.
      
\end{enumerate}

\input{figures/fig1}

Each case leads to a contradiction, implying that every $e \in M \setminus P^*$ connects two middle points of $P^*$.
    
Now, contraction of each $e \in M\setminus P^* $ makes a vertex of degree 4 in $P^*/M$. Contraction of each $e \in M \cap P^* $ makes a vertex of degree 2 and decrements the number of edges in $P^*$. As a result, $P^*/M$ is a graph whose vertices' degrees are 1,2 or 4, $|E(P^*/M)| = |P^*| - |P^* \cap M| $ and $|V_4(P^*/M)| = |M \setminus P^*|$.
Finally, using \hyperref[E-deg4]{Lemma \ref*{E-deg4}} for $P^*/M$ we have
$$ \mu(P^*/M) \geq \frac{|E(P^*/M)| - |V_4(P^*/M)|}{3} = \frac{|P^*| - |P^* \cap M| - |M \setminus P^*|}{3} = \frac{|P^*| - |M|}{3} .$$
Since $P^*/M$ is a subgraph of $G/M$ we have
$$ \mu(G/M) \geq \mu(P^*/M) \geq \frac{|P^*| - |M|}{3} = \frac{\rho(G) - |M|}{3}. $$
\end{myproof}

Using the above results, we compute the approximation factor of \hyperref[alg1]{Algorithm 1}.

\begin{mytheorem}{}\label{ratio-2/3}
	The approximation factor of \hyperref[alg1]{Algorithm 1} is  $ \frac{2}{3} (1 - \epsilon) $, i.e.,
 \begin{equation}\label{approx-M1M2}
     \rho(G) \geq |M_1 \cup M_2| \geq \frac{2}{3} (1 - \epsilon) \rho(G) .
 \end{equation}
\end{mytheorem}

\begin{proof}
    By \hyperref[approx-M1]{Corollary \ref*{approx-M1}} and, \hyperref[mu>(rho-M)/3]{Lemma \ref*{mu>(rho-M)/3}}, we have
 \begin{eqnarray*}
        |M_1 \cup M_2| &=& |M_1| + |M_2| \\
        &\geq& 
        |M_1| + (1-\epsilon)\mu(G/M_1) \\
        &\geq& 
        |M_1| + \frac{1-\epsilon}{3} (\rho(G) - |M_1|) \\
        &\geq&
        \frac{1-\epsilon}{3} \rho(G) + \frac{2}{3} |M_1| \\
        & \geq &
        \frac{1-\epsilon}{3} \rho(G) + \frac{1-\epsilon}{3} \rho(G) =\frac{2}{3} (1- \epsilon ) \rho(G).
    \end{eqnarray*}
    
Since $M_1 \cup M_2$ is a path cover, we have $\rho(G) \geq |M_1\cup M_2| $. Hence, the approximation factor of \hyperref[alg1]{Algorithm 1} is at least $\frac{2}{3} (1- \epsilon)$.

\end{proof}

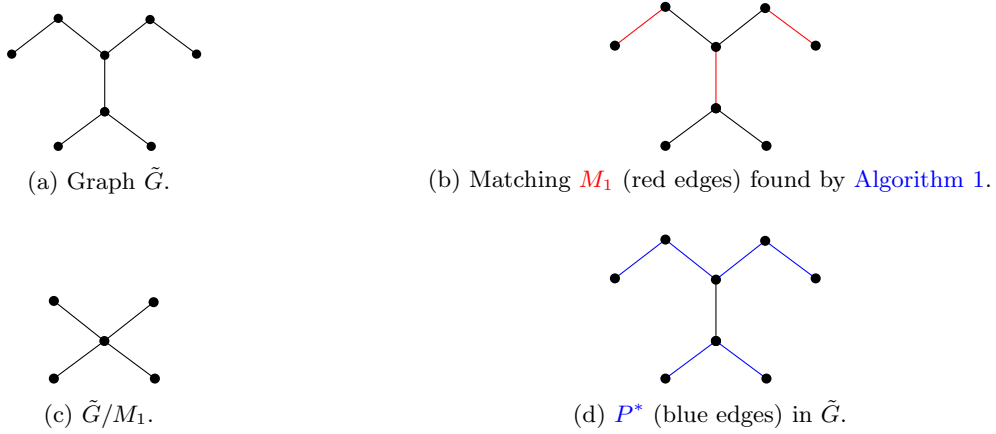
\begin{figure}[h]

\subcaptionbox{Graph $\tilde{G}$. \label{upper-bound-graph}}%
	[.4\linewidth]{\centering
		\begin{tikzpicture}[x=0.75pt,y=0.75pt,yscale=-0.58,xscale=0.58]
				
				\draw    (224.47,124.73) -- (224.47,173.73) ;
				\draw    (184.47,92.73) -- (224.47,124.73) ;
				\draw    (184.47,92.73) -- (144.47,123.73) ;
				\draw    (263.47,93.73) -- (303.47,123.73) ;
				\draw    (224.47,124.73) -- (263.47,93.73) ;
				\draw    (224.47,173.73) -- (184.47,203.73) ;
				\draw    (224.47,173.73) -- (264.47,203.73) ;

			\fill  (224.47,124.73) circle [x radius= 4, y radius= 4]   ;	\fill  (224.47,173.73)  circle [x radius= 4, y radius= 4]   ;	\fill  (184.47,92.73) circle [x radius= 4, y radius= 4]   ;	\fill  (144.47,123.73) circle [x radius= 4, y radius= 4]   ;	\fill  (263.47,93.73) circle [x radius= 4, y radius= 4]   ;	
		\fill (303.47,123.73) circle [x radius= 4, y radius= 4]   ;	
  \fill  (224.47,173.73) circle [x radius= 4, y radius= 4]   ;	
  \fill  (184.47,203.73) circle [x radius= 4, y radius= 4]   ;	
  \fill  (264.47,203.73) circle [x radius= 4, y radius= 4]   ;		
  
			\end{tikzpicture}}
\subcaptionbox{Matching $ \color{red} M_1$ (red edges) found by \hyperref[alg1]{Algorithm 1}. \label{fig-upper-bound-m1}}%
	[.6\linewidth]{\centering
		\begin{tikzpicture}[x=0.75pt,y=0.75pt,yscale=-0.63,xscale=0.63]
				
				\draw [color={rgb, 255:red, 255; green, 0; blue, 0 }  ,draw opacity=1 ]   (224.47,124.73) -- (224.47,173.73) ;
				\draw    (184.47,92.73) -- (224.47,124.73) ;
				\draw [color={rgb, 255:red, 255; green, 2; blue, 2 }  ,draw opacity=1 ]   (184.47,92.73) -- (144.47,123.73) ;
				\draw [color={rgb, 255:red, 255; green, 0; blue, 0 }  ,draw opacity=1 ]   (263.47,93.73) -- (303.47,123.73) ;
				\draw    (224.47,124.73) -- (263.47,93.73) ;
				\draw    (224.47,173.73) -- (184.47,203.73) ;
				\draw    (224.47,173.73) -- (264.47,203.73) ;

						\fill  (224.47,124.73) circle [x radius= 4, y radius= 4]   ;	\fill  (224.47,173.73)  circle [x radius= 4, y radius= 4]   ;	\fill  (184.47,92.73) circle [x radius= 4, y radius= 4]   ;	\fill  (144.47,123.73) circle [x radius= 4, y radius= 4]   ;	\fill  (263.47,93.73) circle [x radius= 4, y radius= 4]   ;	
		\fill (303.47,123.73) circle [x radius= 4, y radius= 4]   ;	
  \fill  (224.47,173.73) circle [x radius= 4, y radius= 4]   ;	
  \fill  (184.47,203.73) circle [x radius= 4, y radius= 4]   ;	
  \fill  (264.47,203.73) circle [x radius= 4, y radius= 4]   ;			
				
			\end{tikzpicture}}
        \hfill
	\par\bigskip

\subcaptionbox{$\tilde{G}/M_1$. \label{upper-bound-tildeG-M1}}%
	[.4\linewidth]{\centering
		\begin{tikzpicture}[x=0.75pt,y=0.75pt,yscale=-0.63,xscale=0.63]
				
				\draw    (184.47,141.73) -- (224.47,173.73) ;
				\draw    (224.47,173.73) -- (263.47,142.73) ;
				\draw    (224.47,173.73) -- (184.47,203.73) ;
				\draw    (224.47,173.73) -- (264.47,203.73) ;

\fill  (184.47,141.73) circle [x radius= 4, y radius= 4]   ;	
  \fill  (224.47,173.73) circle [x radius= 4, y radius= 4]   ;	
  \fill  (263.47,142.73) circle [x radius= 4, y radius= 4]   ;	
  \fill  (264.47,203.73) circle [x radius= 4, y radius= 4]   ;		
\fill   (184.47,203.73) circle [x radius= 4, y radius= 4]   ;				
				
			\end{tikzpicture}}
 \subcaptionbox{$ \color{blue} P^*$ (blue edges) in $\tilde{G}$. \label{fig-upper-bound-mpc}}%
	[.6\linewidth]{\centering
		\begin{tikzpicture}[x=0.75pt,y=0.75pt,yscale=-0.63,xscale=0.63]
				
				\draw    (224.47,124.73) -- (224.47,173.73) ;
				\draw  [color={rgb, 255:red, 0; green, 2; blue, 255 }  ,draw opacity=1 ]    (184.47,92.73) -- (224.47,124.73) ;
				\draw [color={rgb, 255:red, 0; green, 2; blue, 255 }  ,draw opacity=1 ]   (184.47,92.73) -- (144.47,123.73) ;
				\draw [color={rgb, 255:red, 0; green, 0; blue, 255 }  ,draw opacity=1 ]   (263.47,93.73) -- (303.47,123.73) ;
				\draw  [color={rgb, 255:red, 0; green, 2; blue, 255 }  ,draw opacity=1 ]    (224.47,124.73) -- (263.47,93.73) ;
				\draw  [color={rgb, 255:red, 0; green, 2; blue, 255 }  ,draw opacity=1 ]    (224.47,173.73) -- (184.47,203.73) ;
				\draw  [color={rgb, 255:red, 0; green, 2; blue, 255 }  ,draw opacity=1 ]    (224.47,173.73) -- (264.47,203.73) ;

							\fill  (224.47,124.73) circle [x radius= 4, y radius= 4]   ;	\fill  (224.47,173.73)  circle [x radius= 4, y radius= 4]   ;	\fill  (184.47,92.73) circle [x radius= 4, y radius= 4]   ;	\fill  (144.47,123.73) circle [x radius= 4, y radius= 4]   ;	\fill  (263.47,93.73) circle [x radius= 4, y radius= 4]   ;	
		\fill (303.47,123.73) circle [x radius= 4, y radius= 4]   ;	
  \fill  (224.47,173.73) circle [x radius= 4, y radius= 4]   ;	
  \fill  (184.47,203.73) circle [x radius= 4, y radius= 4]   ;	
  \fill  (264.47,203.73) circle [x radius= 4, y radius= 4]   ;		
				
			\end{tikzpicture}}
 
 \caption{
An example of a graph $\tilde{G}$ for which \hyperref[alg1]{Algorithm 1} produces a path cover whose size is $\frac{2}{3}$ times the size of the MPC.
 }
\end{figure}

Now, we show that our analysis of the approximation factor of \hyperref[alg1]{Algorithm 1} is tight.
Consider the graph in  \hyperref[upper-bound-graph]{Figure \ref*{upper-bound-graph}} and denote it by $\tilde{G}$.
If we run \hyperref[alg1]{Algorithm 1} on $\tilde{G}$, then the edges of $M_1$ could be the red edges shown in \hyperref[fig-upper-bound-m1]{Figure \ref*{fig-upper-bound-m1}}.
After contracting $\tilde{G}$ on $M_1$, we have $\tilde{G}/M_1$ shown in \hyperref[upper-bound-tildeG-M1]{Figure \ref*{upper-bound-tildeG-M1}}.
Finally, the second matching $M_2$ found by \hyperref[alg1]{Algorithm 1} in $\tilde{G}/M_1$ contains at most one edge which implies
$|M_1 \cup M_2| \leq 4$.
On the other hand, maximum path cover $P^*$ in $\tilde{G}$ contains $6$ edges shown in  \hyperref[fig-upper-bound-mpc]{Figure \ref*{fig-upper-bound-mpc}}.

As a result, 
 $\dfrac{|M_1 \cup M_2|}{|P^*|} \leq \frac{2}{3}$, so this example and  \hyperref[ratio-2/3]{Theorem \ref*{ratio-2/3}} imply that the approximation factor of \hyperref[alg1]{Algorithm 1} is $\frac{2}{3}-\epsilon$.

    \subsection{Implementation of \hyperref[alg1]{Algorithm 1} in the Semi-Streaming Model}

Now we explain how to implement \hyperref[alg1]{Algorithm 1} in the semi-streaming model. We start with the following theorem by  Fischer, Mitrovic and Uitto \cite{FischerMU22}.

\begin{mytheorem}(Theorem 1.1 in \cite{FischerMU22})
\label{smuw}
Given a graph on $n$ vertices, there is a deterministic
$(1 -\epsilon)$-approximation algorithm for maximum matching that runs
in $\text{poly}(\frac{1}{\epsilon})$ passes in the semi-streaming model. Furthermore, the
algorithm requires  $n \cdot \text{poly}(\frac{1}{\epsilon})$ words of memory.
\end{mytheorem}

To implement \hyperref[alg1]{Algorithm 1} in the semi-streaming model, we proceed as follows: In the first phase, by applying \hyperref[smuw]{Theorem \ref*{smuw}}, we compute a $(1 - \epsilon)$-approximate matching for the graph $ G $, denoted as $M_1$. At the end of this phase, we have the edges of this matching. In the second phase, we again apply \hyperref[smuw]{Theorem \ref*{smuw}} to compute a matching for $G/M_1$ in the streaming model.

During the second phase, when we apply the algorithm of \hyperref[smuw]{Theorem \ref*{smuw}}, while processing each edge $(v_i, v_j)$ in the stream, we follow these rules: If $(v_i, v_j) \in M_1$, we ignore this edge. If $(v_i, v_j) \notin M_1$, but one of $v_i$  or $v_j$  is an endpoint of an edge in $M_1$ (e.g., $v_i, v_k \in M_1 $), then since $(v_i, v_j)$ is contracted, we consider $v_i$ and  $v_j$ as a single vertex, $v_{ij}$. In this case, $ v_k$  is considered adjacent to the new vertex  $v_{ij}$. Consequently, we can compute a $(1 - \epsilon)$-approximation matching for $G/M_1$ in the next $\text{poly}(1/\epsilon)$  passes.

Thus, combining the results of \hyperref[alg1]{Algorithm 1},  \hyperref[ratio-2/3]{Theorem \ref*{ratio-2/3}}, and \hyperref[smuw]{Theorem \ref*{smuw}}, we have the main result of this section:

\begin{mytheorem}{}
\label{main1}
    Given an unweighted graph $G$ on $n$ vertices, there is a deterministic algorithm that returns a $(\frac{2}{3}-\epsilon)$-approximate MPC in the semi-streaming model in $\text{poly}(\frac{1}{\epsilon})$ passes.   
\end{mytheorem}

\subsection{Proof of \hyperref[E-deg4]{Lemma \ref*{E-deg4}}}\label{appendix1}

For simplicity, define 
$$\eta(G) \coloneqq \frac{|E(G)| - |V_4(G)|}{3} .$$
We use a combination of induction and a charging scheme to prove the lemma. Assume 
$$E_{2,4}(G) = \left\{ uv \in E(G) \mid \deg(u)\neq 1, \deg(v) \neq 1 \right\} $$ 
is the set of edges induced on vertices of degree 2 and 4. We use induction on $|E_{2,4}(G)|$.

The base case is when $ E_{2,4}(G) = \emptyset $. In this case, $G$ would be a vertex disjoint union of some $P_2$ (path of length 1), $P_3$ (path of length 2) and $S_5$ (star with 5 vertices and 4 edges). This is because every vertex of degree 2 or 4 could be only incident to vertices of degree 1 which form a connected component $P_3$ or $S_5$ in $G$. Two incident vertices of degree 1 form a connected component $P_2$ in $G$.
It is clear that in each component the inequality holds, and summing them up implies the inequality for $G$.

Suppose the lemma is true for all graphs $G'$ such that $|E_{2,4}(G')| < |E_{2,4}(G)|$. We will prove it for $G$.
First, we change $G$ to obtain $\hat{G}$ with some specific properties and argue that $\mu(\hat{G}) \geq \eta(\hat{G}) $ implies $\mu(G) \geq \eta(G) $. Hence, it is sufficient to prove the lemma for this $\hat{G}$ with some specific properties. 
One important statement here is that it is possible for $\hat{G}$ to contain more edges than $G$, but we construct it such that $|E_{2,4}(\hat{G})| \leq |E_{2,4}(G)|$. As a result, it does not interfere with induction.
After construction of $\hat{G}$, we use a charging scheme to prove the lemma for $\hat{G}$ which implies the lemma for $G$.

For the first part of the proof, we change $G$ to satisfy the following properties and show that it is sufficient to prove the lemma for this updated $G$.
\begin{enumerate}
    
    \item \label{property-1} For each $uv \in E(G)$, either $\deg(u) = 4$ or $\deg(v) = 4$. 
    
    \item \label{property-2} $G$ does not contain parallel edges.
    
    \item \label{property-3} $G$ does not contain a cycle of length three.
    
    \item \label{property-4} If $v$ is a vertex of degree $4$ in $G$ with neighbors $v_1,v_2,v_3$ and $v_4$, then at most one of $v_i$'s have degree $2$.
    
\end{enumerate}
The proof of this part needs a series of updates on $G$ to obtain the desired graph. For this purpose, we introduce some cases and show how to deal with them to obtain $G'$ and replace it by $G$. In each case, we have to check three conditions. First, the vertices of the new graph have degree 1,2 or 4. Since it is easy to check that in some cases, we do not elaborate on that. Second, $\mu(G') \geq \eta(G')$ implies $\mu(G) \geq \eta(G)$. Third, $|E_{2,4}(G')| \leq |E_{2,4}(G)|$.

The order of the cases is important here. Updating $G$ in a case could make another case to hold for $G$. Thus, in each step, we update $G$ according to case $i$ where $i$ is the least possible number that $i$th case holds for $G$. This would prevent us to check additional cases. For instance, when we are dealing with case 2, we assume that case 1 does not hold for $G$.
We proceed with the following cases and show how to update $G$ in each case.
\begin{enumerate}
    \item
    If $G$ contains an edge $uv$ such that $\deg(u) = \deg(v) = 1$, then it is obvious that $u$ and $v$ are not connected to any other vertex of $G$. Let $G'$ be the graph obtained by removing vertices $u$ and $v$ and edge $uv$ from $G$. We have $ \mu(G) = \mu(G') + 1 $ because every matching in $G'$, together with $uv$ forms a matching in $G$. We also have $|E(G)| = |E(G')| + 1$ and $|V_4(G)| = |V_4(G')|$ which conclude $\eta(G) = \eta(G') + \frac{1}{3}$. Hence, having $\mu(G') \geq \eta(G') $ implies 
    $$\mu(G) = \mu(G') + 1 \geq \eta(G') + 1 > \eta(G') + \frac{1}{3} = \eta(G) .$$
    Note that $|E_{2,4}(G')| = |E_{2,4}(G)|$. 

    \item 
    If $G$ contains two parallel edges between $u$ and $v$, then by symmetry between $u$ and $v$, we have three subcases as follows.
    \begin{enumerate}
        \item If $\deg(u) = \deg(v)=2$, then $u$ and $v$ are not connected to any other vertex in $G$. Let $G'$ be the graph obtained by removing $u$ and $v$ and two parallel edges between $u$ and $v$. We have $|V_4(G)| = |V_4(G')| $ and $|E(G)| = |E(G')| + 2 $. As a result, $ \eta(G) = \eta(G') + \frac{2}{3} $. We also have $\mu(G) = \mu(G') + 1 $ because a maximum matching in $G'$, together with edge $uv$ forms a matching in $G$.
        Finally, $\mu(G') \geq \eta(G') $ implies 
        $$\mu(G) = \mu(G') + 1 \geq \eta(G') + 1 > \eta(G') + \frac{2}{3} = \eta(G) .$$ 
        Note that $|E_{2,4}(G')| = |E_{2,4}(G)|-2$, because we removed two edges between $u$ and $v$ from $E_{2,4}(G') $.

        \item If $\deg(u) = 2$ and $\deg(v)=4$, then let $G'$ be the graph obtained by replacing two vertices $u_1$ and $u_2$ instead of $u$ and edges $u_1v$ and $u_2v$ instead of two parallel edges between $u$ and $v$. We have $|V_4(G)| = |V_4(G')| $ and $|E(G)| = |E(G')| $. As a result, $ \eta(G) = \eta(G') $. We also have $\mu(G) = \mu(G') $.
        Finally, $\mu(G') \geq \eta(G') $ implies 
        $$\mu(G) = \mu(G')  \geq \eta(G') = \eta(G) .$$ 
        Note that $|E_{2,4}(G')| = |E_{2,4}(G)|-2$.

        \begin{center}
    \tikzset{every picture/.style={line width=0.75pt}} 

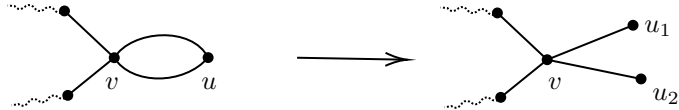
\begin{tikzpicture}[x=0.75pt,y=0.75pt,yscale=-1,xscale=1]

\draw    (213.47,143.07) -- (244.44,143.63) -- (266.47,144.03) ;
\draw [shift={(268.47,144.07)}, rotate = 181.04] [color={rgb, 255:red, 0; green, 0; blue, 0 }  ][line width=0.75]    (10.93,-3.29) .. controls (6.95,-1.4) and (3.31,-0.3) .. (0,0) .. controls (3.31,0.3) and (6.95,1.4) .. (10.93,3.29)   ;
\draw  [fill={rgb, 255:red, 0; green, 0; blue, 0 }  ,fill opacity=1 ] (120,143.07) .. controls (120,144.3) and (121,145.3) .. (122.23,145.3) .. controls (123.47,145.3) and (124.47,144.3) .. (124.47,143.07) .. controls (124.47,141.83) and (123.47,140.83) .. (122.23,140.83) .. controls (121,140.83) and (120,141.83) .. (120,143.07) -- cycle ;
\draw    (97.4,119.53) -- (122.23,143.07) ;
\draw    (122.23,143.07) .. controls (131.07,129) and (161.07,127) .. (169.07,143) ;
\draw    (122.23,143.07) .. controls (130.83,157) and (157.07,157) .. (169.07,143) ;
\draw    (122.23,143.07) -- (99.07,162) ;
\draw  [fill={rgb, 255:red, 0; green, 0; blue, 0 }  ,fill opacity=1 ] (95.17,119.53) .. controls (95.17,120.77) and (96.17,121.77) .. (97.4,121.77) .. controls (98.63,121.77) and (99.63,120.77) .. (99.63,119.53) .. controls (99.63,118.3) and (98.63,117.3) .. (97.4,117.3) .. controls (96.17,117.3) and (95.17,118.3) .. (95.17,119.53) -- cycle ;
\draw  [fill={rgb, 255:red, 0; green, 0; blue, 0 }  ,fill opacity=1 ] (166.83,143) .. controls (166.83,144.23) and (167.83,145.23) .. (169.07,145.23) .. controls (170.3,145.23) and (171.3,144.23) .. (171.3,143) .. controls (171.3,141.77) and (170.3,140.77) .. (169.07,140.77) .. controls (167.83,140.77) and (166.83,141.77) .. (166.83,143) -- cycle ;
\draw  [fill={rgb, 255:red, 0; green, 0; blue, 0 }  ,fill opacity=1 ] (96.83,162) .. controls (96.83,163.23) and (97.83,164.23) .. (99.07,164.23) .. controls (100.3,164.23) and (101.3,163.23) .. (101.3,162) .. controls (101.3,160.77) and (100.3,159.77) .. (99.07,159.77) .. controls (97.83,159.77) and (96.83,160.77) .. (96.83,162) -- cycle ;
\draw  [dash pattern={on 0.75pt off 0.75pt}]  (97.4,119.53) .. controls (95.58,121.03) and (93.92,120.87) .. (92.42,119.05) .. controls (90.93,117.23) and (89.27,117.07) .. (87.45,118.56) .. controls (85.63,120.06) and (83.97,119.9) .. (82.47,118.08) .. controls (80.97,116.26) and (79.31,116.1) .. (77.49,117.59) .. controls (75.67,119.08) and (74.01,118.92) .. (72.52,117.1) -- (69.07,116.77) -- (69.07,116.77) ;
\draw  [dash pattern={on 0.75pt off 0.75pt}]  (99.07,162) .. controls (97.72,163.93) and (96.07,164.23) .. (94.14,162.88) .. controls (92.21,161.53) and (90.57,161.83) .. (89.22,163.76) .. controls (87.87,165.69) and (86.23,165.99) .. (84.3,164.64) .. controls (82.37,163.29) and (80.73,163.59) .. (79.38,165.52) .. controls (78.03,167.45) and (76.39,167.74) .. (74.46,166.39) -- (71.07,167) -- (71.07,167) ;
\draw  [fill={rgb, 255:red, 0; green, 0; blue, 0 }  ,fill opacity=1 ] (336,144.07) .. controls (336,145.3) and (337,146.3) .. (338.23,146.3) .. controls (339.47,146.3) and (340.47,145.3) .. (340.47,144.07) .. controls (340.47,142.83) and (339.47,141.83) .. (338.23,141.83) .. controls (337,141.83) and (336,142.83) .. (336,144.07) -- cycle ;
\draw    (313.4,120.53) -- (338.23,144.07) ;
\draw    (338.23,144.07) -- (315.07,163) ;
\draw    (381.07,126.73) -- (338.23,144.07) ;
\draw    (338.23,144.07) -- (385.07,153.73) ;
\draw  [fill={rgb, 255:red, 0; green, 0; blue, 0 }  ,fill opacity=1 ] (311.17,120.53) .. controls (311.17,121.77) and (312.17,122.77) .. (313.4,122.77) .. controls (314.63,122.77) and (315.63,121.77) .. (315.63,120.53) .. controls (315.63,119.3) and (314.63,118.3) .. (313.4,118.3) .. controls (312.17,118.3) and (311.17,119.3) .. (311.17,120.53) -- cycle ;
\draw  [fill={rgb, 255:red, 0; green, 0; blue, 0 }  ,fill opacity=1 ] (378.83,126.73) .. controls (378.83,127.97) and (379.83,128.97) .. (381.07,128.97) .. controls (382.3,128.97) and (383.3,127.97) .. (383.3,126.73) .. controls (383.3,125.5) and (382.3,124.5) .. (381.07,124.5) .. controls (379.83,124.5) and (378.83,125.5) .. (378.83,126.73) -- cycle ;
\draw  [fill={rgb, 255:red, 0; green, 0; blue, 0 }  ,fill opacity=1 ] (382.83,153.73) .. controls (382.83,154.97) and (383.83,155.97) .. (385.07,155.97) .. controls (386.3,155.97) and (387.3,154.97) .. (387.3,153.73) .. controls (387.3,152.5) and (386.3,151.5) .. (385.07,151.5) .. controls (383.83,151.5) and (382.83,152.5) .. (382.83,153.73) -- cycle ;
\draw  [fill={rgb, 255:red, 0; green, 0; blue, 0 }  ,fill opacity=1 ] (312.83,163) .. controls (312.83,164.23) and (313.83,165.23) .. (315.07,165.23) .. controls (316.3,165.23) and (317.3,164.23) .. (317.3,163) .. controls (317.3,161.77) and (316.3,160.77) .. (315.07,160.77) .. controls (313.83,160.77) and (312.83,161.77) .. (312.83,163) -- cycle ;
\draw  [dash pattern={on 0.75pt off 0.75pt}]  (313.4,120.53) .. controls (311.58,122.03) and (309.92,121.87) .. (308.42,120.05) .. controls (306.93,118.23) and (305.27,118.07) .. (303.45,119.56) .. controls (301.63,121.06) and (299.97,120.9) .. (298.47,119.08) .. controls (296.97,117.26) and (295.31,117.1) .. (293.49,118.59) .. controls (291.67,120.08) and (290.01,119.92) .. (288.52,118.1) -- (285.07,117.77) -- (285.07,117.77) ;
\draw  [dash pattern={on 0.75pt off 0.75pt}]  (315.07,163) .. controls (313.72,164.93) and (312.07,165.23) .. (310.14,163.88) .. controls (308.21,162.53) and (306.57,162.83) .. (305.22,164.76) .. controls (303.87,166.69) and (302.23,166.99) .. (300.3,165.64) .. controls (298.37,164.29) and (296.73,164.59) .. (295.38,166.52) .. controls (294.03,168.45) and (292.39,168.74) .. (290.46,167.39) -- (287.07,168) -- (287.07,168) ;

\draw (116.23,151.7) node [anchor=north west][inner sep=0.75pt]    {$v$};
\draw (164.23,151.7) node [anchor=north west][inner sep=0.75pt]    {$u$};
\draw (337.23,151.7) node [anchor=north west][inner sep=0.75pt]    {$v$};
\draw (389.3,157.13) node [anchor=north west][inner sep=0.75pt]    {$u_{2}$};
\draw (385.3,123.13) node [anchor=north west][inner sep=0.75pt]    {$u_{1}$};

\end{tikzpicture}
\end{center}

        \item If $\deg(u) = \deg(v) = 4 $, then let $G'$ be the graph obtained by removing these two parallel edges from  $G$. We have $|V_4(G)| = |V_4(G')| + 2 $ and $|E(G)| = |E(G')| + 2 $. As a result, $ \eta(G) = \eta(G') $. We also have $\mu(G) \geq \mu(G') $, because a maximum matching in $G'$ is also a matching in $G$.
        Finally, $\mu(G') \geq \eta(G') $ implies 
        $$\mu(G) \geq \mu(G') \geq \eta(G') = \eta(G) .$$
        Note that $|E_{2,4}(G')| = |E_{2,4}(G)|-2$.
        
        \input{figures/fig-appendixA-parallel-c}
        
    \end{enumerate}

    \item
    If $G$ contains an edge $uv$ such that $\deg(u) = \deg(v) = 2 $. If there are 2 parallel edges between $u$ and $v$, then this case has been covered in Case 2(a). So, assume that $uu_1, vv_1 \in E(G)$ are edges different from $uv$ in $G$. If $u_1 \neq v_1$, then $u_1uvv_1$ is a path of length three. Otherwise, $u_1=v_1$ which means $u_1uv$ is a cycle of length three. We explain each case as follows.
    \begin{enumerate}
        \item $u_1=v_1$. In this case let $G'$ be the graph obtained by removing vertices $u$ and $v$ and edges $uv,u_1u$ and $vv_1$ from $G$. We have $|E(G)| = |E(G')| + 3 $. If $\deg(u_1) = 4$, then $|V_4(G)| = |V_4(G')| + 1$, and if $\deg(v_1) = 2$, then $|V_4(G)| = |V_4(G')| $. As a result, in both cases we have $\eta(G') + 1 \geq \eta(G) $.
        Every matching in $G'$, together with $uv$ forms a matching in $G$ which implies $\mu(G) \geq \mu(G') + 1$. Finally, $\mu(G') \geq \eta(G') $ implies 
        $$\mu(G) \geq \mu(G') + 1 \geq \eta(G') + 1 \geq \eta(G) .$$ 
        Note that $|E_{2,4}(G')| = |E_{2,4}(G)|-3$.

        \begin{center}
    \tikzset{every picture/.style={line width=0.75pt}} 

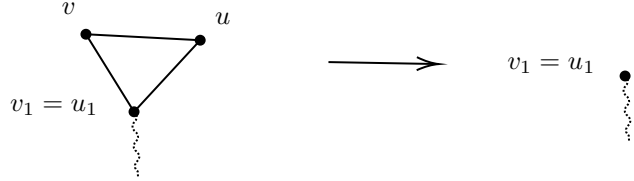
\begin{tikzpicture}[x=0.75pt,y=0.75pt,yscale=-1,xscale=1]

\draw    (134.47,137.07) -- (110.47,98.07) ;
\draw  [fill={rgb, 255:red, 0; green, 0; blue, 0 }  ,fill opacity=1 ] (108.23,98.07) .. controls (108.23,99.3) and (109.23,100.3) .. (110.47,100.3) .. controls (111.7,100.3) and (112.7,99.3) .. (112.7,98.07) .. controls (112.7,96.83) and (111.7,95.83) .. (110.47,95.83) .. controls (109.23,95.83) and (108.23,96.83) .. (108.23,98.07) -- cycle ;
\draw  [fill={rgb, 255:red, 0; green, 0; blue, 0 }  ,fill opacity=1 ] (132.23,137.07) .. controls (132.23,138.3) and (133.23,139.3) .. (134.47,139.3) .. controls (135.7,139.3) and (136.7,138.3) .. (136.7,137.07) .. controls (136.7,135.83) and (135.7,134.83) .. (134.47,134.83) .. controls (133.23,134.83) and (132.23,135.83) .. (132.23,137.07) -- cycle ;
\draw  [fill={rgb, 255:red, 0; green, 0; blue, 0 }  ,fill opacity=1 ] (165.23,101.07) .. controls (165.23,102.3) and (166.23,103.3) .. (167.47,103.3) .. controls (168.7,103.3) and (169.7,102.3) .. (169.7,101.07) .. controls (169.7,99.83) and (168.7,98.83) .. (167.47,98.83) .. controls (166.23,98.83) and (165.23,99.83) .. (165.23,101.07) -- cycle ;
\draw    (231.47,112.07) -- (284.47,113.03) ;
\draw [shift={(286.47,113.07)}, rotate = 181.04] [color={rgb, 255:red, 0; green, 0; blue, 0 }  ][line width=0.75]    (10.93,-3.29) .. controls (6.95,-1.4) and (3.31,-0.3) .. (0,0) .. controls (3.31,0.3) and (6.95,1.4) .. (10.93,3.29)   ;
\draw [color={rgb, 255:red, 0; green, 0; blue, 0 }  ,draw opacity=1 ]   (134.47,137.07) -- (167.47,101.07) ;
\draw    (110.47,98.07) -- (167.47,101.07) ;
\draw  [dash pattern={on 0.75pt off 0.75pt}]  (134.47,137.07) .. controls (136.23,138.63) and (136.33,140.29) .. (134.77,142.06) .. controls (133.2,143.83) and (133.3,145.49) .. (135.07,147.05) .. controls (136.84,148.61) and (136.94,150.27) .. (135.37,152.04) .. controls (133.81,153.81) and (133.91,155.47) .. (135.68,157.03) .. controls (137.45,158.59) and (137.55,160.25) .. (135.98,162.02) .. controls (134.41,163.79) and (134.51,165.45) .. (136.28,167.01) -- (136.47,170.07) -- (136.47,170.07) ;
\draw  [fill={rgb, 255:red, 0; green, 0; blue, 0 }  ,fill opacity=1 ] (377.23,119.07) .. controls (377.23,120.3) and (378.23,121.3) .. (379.47,121.3) .. controls (380.7,121.3) and (381.7,120.3) .. (381.7,119.07) .. controls (381.7,117.83) and (380.7,116.83) .. (379.47,116.83) .. controls (378.23,116.83) and (377.23,117.83) .. (377.23,119.07) -- cycle ;
\draw  [dash pattern={on 0.75pt off 0.75pt}]  (379.47,119.07) .. controls (381.23,120.63) and (381.33,122.29) .. (379.77,124.06) .. controls (378.2,125.83) and (378.3,127.49) .. (380.07,129.05) .. controls (381.84,130.61) and (381.94,132.27) .. (380.37,134.04) .. controls (378.81,135.81) and (378.91,137.47) .. (380.68,139.03) .. controls (382.45,140.59) and (382.55,142.25) .. (380.98,144.02) .. controls (379.41,145.79) and (379.51,147.45) .. (381.28,149.01) -- (381.47,152.07) -- (381.47,152.07) ;

\draw (173.47,85.7) node [anchor=north west][inner sep=0.75pt]    {$u$};
\draw (97,80.4) node [anchor=north west][inner sep=0.75pt]    {$v$};
\draw (71.23,129.47) node [anchor=north west][inner sep=0.75pt]    {$v_{1} =u_{1}$};
\draw (319.23,108.47) node [anchor=north west][inner sep=0.75pt]    {$v_{1} =u_{1}$};

\end{tikzpicture}
\end{center}

        \item $u_1 \neq v_1$. In this case let $G'$ be the graph obtained by removing $u$ and $v$ and edges $uv,u_1u$ and $vv_1$ from $G$ and adding a new edge $u_1v_1$ between $u_1$ and $v_1$ (Note that there might be an edge between $u_1$ and $v_1$ already, but we add this edge to prevent creating vertices of degree $3$ in $G'$). We have $|V_4(G)| = |V_4(G')| $ and $|E(G)| = |E(G')| + 2 $. As a result, $ \eta(G) = \eta(G') + \frac{2}{3} $.
        Now, consider a maximum matching $M'$ in $G'$. If $M'$ does not contain the new edge $u_1v_1$, then add $uv$ to $M'$ to get a matching in $G$ of size $\mu(G') + 1$. If $M'$ contains $u_1v_1$, then remove it from $M'$ and add two edges $u_1u$ and $vv_1$ to obtain a matching of size $\mu(G') + 1$ in $G$. Hence, in both cases we have $\mu(G) \geq \mu(G') + 1 $.
        Finally, $\mu(G') \geq \eta(G') $ implies 
        $$\mu(G) \geq \mu(G') + 1 \geq \eta(G') + 1 > \eta(G') + \frac{2}{3} = \eta(G) .$$ 
        Note that $|E_{2,4}(G')| \leq |E_{2,4}(G)|$, because we removed at least one edge $ uv $ from $ E_{2,4}(G) $ and added at most one edge $ u_1v_1 $ to it.

        \begin{center}
    
\tikzset{every picture/.style={line width=0.75pt}} 

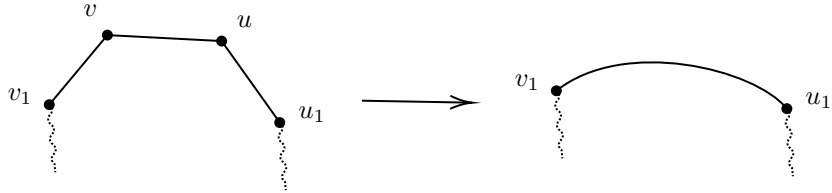
\begin{tikzpicture}[x=0.75pt,y=0.75pt,yscale=-1,xscale=1]

\draw    (81.47,133.07) -- (110.47,98.07) ;
\draw  [fill={rgb, 255:red, 0; green, 0; blue, 0 }  ,fill opacity=1 ] (108.23,98.07) .. controls (108.23,99.3) and (109.23,100.3) .. (110.47,100.3) .. controls (111.7,100.3) and (112.7,99.3) .. (112.7,98.07) .. controls (112.7,96.83) and (111.7,95.83) .. (110.47,95.83) .. controls (109.23,95.83) and (108.23,96.83) .. (108.23,98.07) -- cycle ;
\draw  [fill={rgb, 255:red, 0; green, 0; blue, 0 }  ,fill opacity=1 ] (79.23,133.07) .. controls (79.23,134.3) and (80.23,135.3) .. (81.47,135.3) .. controls (82.7,135.3) and (83.7,134.3) .. (83.7,133.07) .. controls (83.7,131.83) and (82.7,130.83) .. (81.47,130.83) .. controls (80.23,130.83) and (79.23,131.83) .. (79.23,133.07) -- cycle ;
\draw  [fill={rgb, 255:red, 0; green, 0; blue, 0 }  ,fill opacity=1 ] (165.23,101.07) .. controls (165.23,102.3) and (166.23,103.3) .. (167.47,103.3) .. controls (168.7,103.3) and (169.7,102.3) .. (169.7,101.07) .. controls (169.7,99.83) and (168.7,98.83) .. (167.47,98.83) .. controls (166.23,98.83) and (165.23,99.83) .. (165.23,101.07) -- cycle ;
\draw    (237.47,131.07) -- (268.44,131.63) -- (290.47,132.03) ;
\draw [shift={(292.47,132.07)}, rotate = 181.04] [color={rgb, 255:red, 0; green, 0; blue, 0 }  ][line width=0.75]    (10.93,-3.29) .. controls (6.95,-1.4) and (3.31,-0.3) .. (0,0) .. controls (3.31,0.3) and (6.95,1.4) .. (10.93,3.29)   ;
\draw [color={rgb, 255:red, 0; green, 0; blue, 0 }  ,draw opacity=1 ]   (196.47,142.07) -- (167.47,101.07) ;
\draw    (110.47,98.07) -- (167.47,101.07) ;
\draw  [fill={rgb, 255:red, 0; green, 0; blue, 0 }  ,fill opacity=1 ] (194.23,142.07) .. controls (194.23,143.3) and (195.23,144.3) .. (196.47,144.3) .. controls (197.7,144.3) and (198.7,143.3) .. (198.7,142.07) .. controls (198.7,140.83) and (197.7,139.83) .. (196.47,139.83) .. controls (195.23,139.83) and (194.23,140.83) .. (194.23,142.07) -- cycle ;
\draw  [fill={rgb, 255:red, 0; green, 0; blue, 0 }  ,fill opacity=1 ] (332.23,126.07) .. controls (332.23,127.3) and (333.23,128.3) .. (334.47,128.3) .. controls (335.7,128.3) and (336.7,127.3) .. (336.7,126.07) .. controls (336.7,124.83) and (335.7,123.83) .. (334.47,123.83) .. controls (333.23,123.83) and (332.23,124.83) .. (332.23,126.07) -- cycle ;
\draw  [fill={rgb, 255:red, 0; green, 0; blue, 0 }  ,fill opacity=1 ] (447.23,135.07) .. controls (447.23,136.3) and (448.23,137.3) .. (449.47,137.3) .. controls (450.7,137.3) and (451.7,136.3) .. (451.7,135.07) .. controls (451.7,133.83) and (450.7,132.83) .. (449.47,132.83) .. controls (448.23,132.83) and (447.23,133.83) .. (447.23,135.07) -- cycle ;
\draw    (334.47,126.07) .. controls (368.23,99.07) and (433.47,115.07) .. (449.47,135.07) ;
\draw  [dash pattern={on 0.75pt off 0.75pt}]  (81.47,133.07) .. controls (83.28,134.58) and (83.42,136.24) .. (81.91,138.05) .. controls (80.4,139.86) and (80.54,141.52) .. (82.35,143.03) .. controls (84.16,144.54) and (84.3,146.2) .. (82.79,148.01) .. controls (81.27,149.82) and (81.41,151.48) .. (83.22,152.99) .. controls (85.03,154.5) and (85.17,156.16) .. (83.66,157.97) .. controls (82.15,159.78) and (82.29,161.44) .. (84.1,162.95) -- (84.47,167.07) -- (84.47,167.07) ;
\draw  [dash pattern={on 0.75pt off 0.75pt}]  (196.47,142.07) .. controls (198.28,143.58) and (198.42,145.24) .. (196.91,147.05) .. controls (195.4,148.86) and (195.54,150.52) .. (197.35,152.03) .. controls (199.16,153.54) and (199.3,155.2) .. (197.79,157.01) .. controls (196.27,158.82) and (196.41,160.48) .. (198.22,161.99) .. controls (200.03,163.5) and (200.17,165.16) .. (198.66,166.97) .. controls (197.15,168.78) and (197.29,170.44) .. (199.1,171.95) -- (199.47,176.07) -- (199.47,176.07) ;
\draw  [dash pattern={on 0.75pt off 0.75pt}]  (334.47,126.07) .. controls (336.28,127.58) and (336.42,129.24) .. (334.91,131.05) .. controls (333.4,132.86) and (333.54,134.52) .. (335.35,136.03) .. controls (337.16,137.54) and (337.3,139.2) .. (335.79,141.01) .. controls (334.27,142.82) and (334.41,144.48) .. (336.22,145.99) .. controls (338.03,147.5) and (338.17,149.16) .. (336.66,150.97) .. controls (335.15,152.78) and (335.29,154.44) .. (337.1,155.95) -- (337.47,160.07) -- (337.47,160.07) ;
\draw  [dash pattern={on 0.75pt off 0.75pt}]  (449.47,135.07) .. controls (451.28,136.58) and (451.42,138.24) .. (449.91,140.05) .. controls (448.4,141.86) and (448.54,143.52) .. (450.35,145.03) .. controls (452.16,146.54) and (452.3,148.2) .. (450.79,150.01) .. controls (449.27,151.82) and (449.41,153.48) .. (451.22,154.99) .. controls (453.03,156.5) and (453.17,158.16) .. (451.66,159.97) .. controls (450.15,161.78) and (450.29,163.44) .. (452.1,164.95) -- (452.47,169.07) -- (452.47,169.07) ;

\draw (173.47,86.7) node [anchor=north west][inner sep=0.75pt]    {$u$};
\draw (97,81.4) node [anchor=north west][inner sep=0.75pt]    {$v$};
\draw (59.23,123.47) node [anchor=north west][inner sep=0.75pt]    {$v_{1}$};
\draw (204.23,132.47) node [anchor=north west][inner sep=0.75pt]    {$u_{1}$};
\draw (312.23,116.47) node [anchor=north west][inner sep=0.75pt]    {$v_{1}$};
\draw (457.23,125.47) node [anchor=north west][inner sep=0.75pt]    {$u_{1}$};

\end{tikzpicture}

\end{center}
        
    \end{enumerate}

    \item 
    Suppose $G$ contains a cycle of length three like $uvw$. Since we have dealt with case 3, at most one of $u,v$ and $w$ has degree 2. Hence, there are two sub cases as follows.
    \begin{enumerate}
        
        \item If $\deg(u) = \deg(v) = \deg(w) = 4$, then let $G'$ be the graph obtained by removing edges $uv$,$vw$ and $wu$ from $G$. We have $|V_4(G)| = |V_4(G')| + 3 $ and $|E(G)| = |E(G')| + 3 $. As a result, $ \eta(G) = \eta(G') $. We also have $\mu(G) \geq \mu(G') $, because a maximum matching in $G'$ is also a matching in $G$.
        Finally, $\mu(G') \geq \eta(G') $ implies 
        $$\mu(G) \geq \mu(G') \geq \eta(G') = \eta(G) .$$ 
        Note that $|E_{2,4}(G')| = |E_{2,4}(G)|-3$. 

        \input{figures/fig-appendixA-cycle-a}
        
        \item If $\deg(u) = \deg(v) = 4 $ and $\deg(w) = 2$, then let $G'$ be the graph obtained by removing edges $uv$ and $uw$ from $G$, then add a new vertex $x$ and an edge $xv$ in $G$. We have $|V_4(G)| = |V_4(G')| + 1 $ and $|E(G)| = |E(G')| + 1 $. As a result, $ \eta(G) = \eta(G') $. We also have $\mu(G) \geq \mu(G') $ because if $M'$ is a maximum matching in $G'$ containing the new edge $xv$, we can replace this edge with $wv$ to obtain a matching of the same size in $G$.
        Finally, $\mu(G') \geq \eta(G') $ implies 
        $$\mu(G) \geq \mu(G') \geq \eta(G') = \eta(G) .$$ 
        Note that $|E_{2,4}(G')| = |E_{2,4}(G)|-3$, because we removed three edges $uv$ and $uw$ and $uw$ from $E_{2,4}(G)$. Also, the degree of $w$ and the added vertex $x$ is 1 in $G'$.

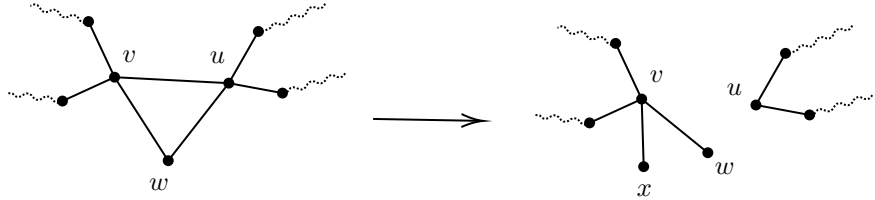
        \input{figures/fig-appendixA-cycle-b}
        
    \end{enumerate}

    \item 
    Suppose $G$ contains a vertex $v$ of degree $4$ with neighbors $v_1,v_2,v_3$ and $v_4$ such that $\deg(v_i) = 2$ for at least two distinct $1 \leq i \leq 4$. Let's $\deg(v_1) = \deg(v_2) = 2$. Since we have handled parallel edges in case 2, we can assume that $v_i$'s are distinct. 
    Let $G'$ be the graph obtained by removing $v$ and edges $vv_1$, $vv_2$, $vv_3$ and $vv_4$ from $G$, then merge vertices $v_1$ and $v_3$ together to obtain a new vertex $v_{13}$ with degree equal to $\deg(v_3)$, and finally merge vertices $v_2$ and $v_4$ together to obtain a new vertex $v_{24}$ with degree equal to $\deg(v_4)$ in $G$ (If $v_1$ and $v_2$ had degree more than 2, this statement could be false and the new graph $G'$ could contain a vertex of degree more than $4$).
    
    Note that since we have dealt with case 4 ($G$ has no cycle of length $3$), this merging does not create loops in the new graph $G'$. More specifically, if $v_1w$ is the edge incident to $v_1$ other than $vv_1$, and $N(v_3)$ is the set of incident edges to $v_3$ excluding $vv_3$, then we have $v_1w \not \in N(v_3)$.

    \input{figures/fig-appendixA-deg4}
    
    We have $|V_4(G)| = |V_4(G')| + 1 $ and $|E(G)| = |E(G')| + 4 $. As a result, $ \eta(G) = \eta(G') + 1 $.
    Now, consider a maximum matching $M'$ in $G'$. The matching $M'$ can not contain both $v_{13}w$ and an edge of $N(v_3)$ in $G'$. If $M'$  does not contain $v_{13}w$, then $M'$ together with $uv_1$ is a matching in $G$. If $M'$ does not contain any edge in $N(v_3)$, then $M'$ together with $uv_3$ is a matching in $G$. Hence, in both cases, we have a matching of size $\mu(G') + 1$ in $G$. Thus, $\mu(G) \geq \mu(G') + 1 $.
    Finally, $\mu(G') \geq \eta(G') $ implies 
    $$\mu(G) \geq \mu(G') + 1 \geq \eta(G') + 1 = \eta(G) .$$
    
    Note that $|E_{2,4}(G')| \leq |E_{2,4}(G)|$, because we have removed four edges from $E_{2,4}(G)$ and merging $v_1$ and $v_3$ which creates a vertex of degree $\deg(v_3)$ in $G'$, does not affect $|E_{2,4}(G')|$. The same reasoning is true for merging $v_2$ and $v_4$.
    
\end{enumerate}
Let $\Phi(G)=|E|+|\{uv  \mid  \text{there is parallel edge between} \ u \ \text{and} \  v\}|$.
In each update, $\Phi(G)$ decreases.
Hence, after a finite number of steps, the process terminates and we obtain a $G$ that none of the cases above hold for it.

Now, we do a final update on $G$ to obtain the desired $\hat{G}$ which satisfies all the properties.
If $G$ contains an edge $uv$ such that $\deg(u) = 1$ and $\deg(v) = 2$, then construct $G'$ by adding two new vertices $v_1$ and $v_2$ to $G$ and two edges $v_1v$ and $v_2v$. We have $|V_4(G)| = |V_4(G')| - 1 $ and $|E(G)| = |E(G')| - 2 $. As a result, $ \eta(G) = \eta(G') - \frac{1}{3} $. We also have $\mu(G) \geq \mu(G') $ because if a maximum matching in $G'$ contains $v_1v$ or $v_2v$, we can replace it with $uv$ to obtain a matching in $G$.
Finally, $\mu(G') \geq \eta(G') $ implies 
$$\mu(G) \geq \mu(G') \geq \eta(G') = \eta(G) + \frac{1}{3} \geq \eta(G) .$$
Note that, in this case, the number of edges in $G'$ is greater than in $G$, but we still have $|E_{2,4}(G')| \leq |E_{2,4}(G)|$   because $v_1v,v_2v \not\in E_{2,4}(G')$ and the degree of $v$ changes from $2$ to $4$, which does not affect $E_{2,4}(G)$. Hence, this process does not interfere with the induction on $|E_{2,4}(G)|$.

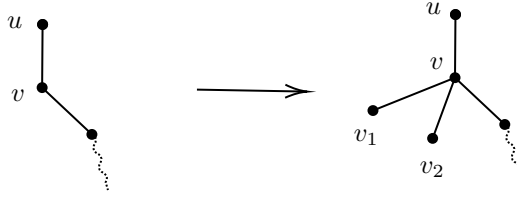
\begin{center}
    \tikzset{every picture/.style={line width=0.75pt}} 

\begin{tikzpicture}[x=0.75pt,y=0.75pt,yscale=-1,xscale=1]

\draw    (199.47,144.07) -- (230.44,144.63) -- (252.47,145.03) ;
\draw [shift={(254.47,145.07)}, rotate = 181.04] [color={rgb, 255:red, 0; green, 0; blue, 0 }  ][line width=0.75]    (10.93,-3.29) .. controls (6.95,-1.4) and (3.31,-0.3) .. (0,0) .. controls (3.31,0.3) and (6.95,1.4) .. (10.93,3.29)   ;
\draw  [fill={rgb, 255:red, 0; green, 0; blue, 0 }  ,fill opacity=1 ] (120,143.07) .. controls (120,144.3) and (121,145.3) .. (122.23,145.3) .. controls (123.47,145.3) and (124.47,144.3) .. (124.47,143.07) .. controls (124.47,141.83) and (123.47,140.83) .. (122.23,140.83) .. controls (121,140.83) and (120,141.83) .. (120,143.07) -- cycle ;
\draw  [dash pattern={on 0.75pt off 0.75pt}]  (147.07,166.6) .. controls (149.13,167.75) and (149.59,169.35) .. (148.44,171.41) .. controls (147.29,173.47) and (147.75,175.07) .. (149.81,176.22) .. controls (151.87,177.36) and (152.33,178.96) .. (151.19,181.02) .. controls (150.04,183.08) and (150.5,184.68) .. (152.56,185.83) .. controls (154.62,186.98) and (155.08,188.58) .. (153.93,190.64) -- (155.07,194.6) -- (155.07,194.6) ;
\draw    (122.47,109.07) -- (122.23,143.07) ;
\draw    (122.23,143.07) -- (147.07,166.6) ;
\draw  [fill={rgb, 255:red, 0; green, 0; blue, 0 }  ,fill opacity=1 ] (120.23,111.3) .. controls (120.23,112.53) and (121.23,113.53) .. (122.47,113.53) .. controls (123.7,113.53) and (124.7,112.53) .. (124.7,111.3) .. controls (124.7,110.07) and (123.7,109.07) .. (122.47,109.07) .. controls (121.23,109.07) and (120.23,110.07) .. (120.23,111.3) -- cycle ;
\draw  [fill={rgb, 255:red, 0; green, 0; blue, 0 }  ,fill opacity=1 ] (120.23,111.3) .. controls (120.23,112.53) and (121.23,113.53) .. (122.47,113.53) .. controls (123.7,113.53) and (124.7,112.53) .. (124.7,111.3) .. controls (124.7,110.07) and (123.7,109.07) .. (122.47,109.07) .. controls (121.23,109.07) and (120.23,110.07) .. (120.23,111.3) -- cycle ;
\draw  [fill={rgb, 255:red, 0; green, 0; blue, 0 }  ,fill opacity=1 ] (326,138.07) .. controls (326,139.3) and (327,140.3) .. (328.23,140.3) .. controls (329.47,140.3) and (330.47,139.3) .. (330.47,138.07) .. controls (330.47,136.83) and (329.47,135.83) .. (328.23,135.83) .. controls (327,135.83) and (326,136.83) .. (326,138.07) -- cycle ;
\draw  [dash pattern={on 0.75pt off 0.75pt}]  (353.07,161.6) .. controls (355.13,162.75) and (355.59,164.35) .. (354.44,166.41) .. controls (353.29,168.47) and (353.75,170.07) .. (355.81,171.22) .. controls (357.87,172.36) and (358.33,173.96) .. (357.19,176.02) .. controls (356.04,178.08) and (356.5,179.68) .. (358.56,180.83) .. controls (360.62,181.98) and (361.08,183.58) .. (359.93,185.64) -- (361.07,189.6) -- (361.07,189.6) ;
\draw    (328.47,104.07) -- (328.23,138.07) ;
\draw    (328.23,138.07) -- (353.07,161.6) ;
\draw  [fill={rgb, 255:red, 0; green, 0; blue, 0 }  ,fill opacity=1 ] (326.23,106.3) .. controls (326.23,107.53) and (327.23,108.53) .. (328.47,108.53) .. controls (329.7,108.53) and (330.7,107.53) .. (330.7,106.3) .. controls (330.7,105.07) and (329.7,104.07) .. (328.47,104.07) .. controls (327.23,104.07) and (326.23,105.07) .. (326.23,106.3) -- cycle ;
\draw  [fill={rgb, 255:red, 0; green, 0; blue, 0 }  ,fill opacity=1 ] (326.23,106.3) .. controls (326.23,107.53) and (327.23,108.53) .. (328.47,108.53) .. controls (329.7,108.53) and (330.7,107.53) .. (330.7,106.3) .. controls (330.7,105.07) and (329.7,104.07) .. (328.47,104.07) .. controls (327.23,104.07) and (326.23,105.07) .. (326.23,106.3) -- cycle ;
\draw    (287.3,154.6) -- (328.23,138.07) ;
\draw    (317.07,168.6) -- (328.23,138.07) ;
\draw  [fill={rgb, 255:red, 0; green, 0; blue, 0 }  ,fill opacity=1 ] (350.83,161.6) .. controls (350.83,162.83) and (351.83,163.83) .. (353.07,163.83) .. controls (354.3,163.83) and (355.3,162.83) .. (355.3,161.6) .. controls (355.3,160.37) and (354.3,159.37) .. (353.07,159.37) .. controls (351.83,159.37) and (350.83,160.37) .. (350.83,161.6) -- cycle ;
\draw  [fill={rgb, 255:red, 0; green, 0; blue, 0 }  ,fill opacity=1 ] (314.83,168.6) .. controls (314.83,169.83) and (315.83,170.83) .. (317.07,170.83) .. controls (318.3,170.83) and (319.3,169.83) .. (319.3,168.6) .. controls (319.3,167.37) and (318.3,166.37) .. (317.07,166.37) .. controls (315.83,166.37) and (314.83,167.37) .. (314.83,168.6) -- cycle ;
\draw  [fill={rgb, 255:red, 0; green, 0; blue, 0 }  ,fill opacity=1 ] (285.07,154.6) .. controls (285.07,155.83) and (286.07,156.83) .. (287.3,156.83) .. controls (288.53,156.83) and (289.53,155.83) .. (289.53,154.6) .. controls (289.53,153.37) and (288.53,152.37) .. (287.3,152.37) .. controls (286.07,152.37) and (285.07,153.37) .. (285.07,154.6) -- cycle ;
\draw  [fill={rgb, 255:red, 0; green, 0; blue, 0 }  ,fill opacity=1 ] (144.83,166.6) .. controls (144.83,167.83) and (145.83,168.83) .. (147.07,168.83) .. controls (148.3,168.83) and (149.3,167.83) .. (149.3,166.6) .. controls (149.3,165.37) and (148.3,164.37) .. (147.07,164.37) .. controls (145.83,164.37) and (144.83,165.37) .. (144.83,166.6) -- cycle ;

\draw (105.23,142.7) node [anchor=north west][inner sep=0.75pt]    {$v$};
\draw (103.23,105.7) node [anchor=north west][inner sep=0.75pt]    {$u$};
\draw (314.23,126.7) node [anchor=north west][inner sep=0.75pt]    {$v$};
\draw (312.23,98.7) node [anchor=north west][inner sep=0.75pt]    {$u$};
\draw (276.23,162.7) node [anchor=north west][inner sep=0.75pt]    {$v_{1}$};
\draw (309.23,178.7) node [anchor=north west][inner sep=0.75pt]    {$v_{2}$};

\end{tikzpicture}
\end{center}

It is easy to see that this update does not cause any of the five cases above to hold for $G$. So, we can finally deal with all the edges $uv$ such that $\deg(u) = 1$ and $\deg(v) = 2$ to obtain the final $G$ that does not satisfy any of the cases above and the parameter $|E_{2,4}(G)|$ has not been increased during these updates.

Note that the final $G$ satisfies the desired properties. Case 2,4 and 5 are equivalent to \hyperref[property-2]{property \ref*{property-2}} ,  \hyperref[property-3]{property \ref*{property-3}} and  \hyperref[property-4]{property \ref*{property-4}}, respectively.
Cases 1,3 and the last update after all five cases on $G$ imply \hyperref[property-1]{property \ref*{property-1}} for $G$.
Also, note that proving $\mu(G) \geq \eta(G)$ for updated $G$ implies this inequality for the original $G$. Hence, we are done with the first part of the proof.

Now, for the rest of the proof, we assume that $G$ has mentioned properties, and we prove $\mu(G) \geq \eta(G)$.

According to \hyperref[property-1]{property \ref*{property-1}}, there are three types of edges in $G$. Let's call edges $uv$ with $\deg(u) = 1$ and $\deg(v) = 4$ \textit{type 1}, edges $uv$ with $\deg(u) = 2$ and $\deg(v) = 4$ \textit{type 2}, and edges $uv$ with $\deg(u) = \deg(v) =  4$ \textit{type 4} (there is no edge defined as type 3). Assume that the number of edges of types 1, 2, and 4 are $a$, $b$, and $c$, respectively. Clearly,
$|E(G)| = a+b+c$. Since each of type 1 and 2 edges have one incident vertex of degree 4 and type 4 edges have two incident vertex of degree 4, then by a simple double counting, we have $|V_4(G)| = \frac{a + b + 2c}{4}$. Hence,
\begin{eqnarray*}
    \mu(G) \geq \eta(G) &\iff& \mu(G) \geq \frac{(a+b+c) - \frac{a + b + 2c}{4}}{3} \\
    &\iff& \mu(G) \geq \frac{3a + 3b + 2c}{12} \\
    &\iff& 12\mu(G) \geq 3a + 3b + 2c
\end{eqnarray*}
Now, we introduce a charging scheme to prove the above inequality. 
Assume $\mathcal{M}$ is the set of all maximum matchings in $G$ that contain the maximum number of edges of type 1. Fix a matching  $M \in \mathcal{M}$. 
For each $e \in M$, we are going to charge at most 12 units to the edges in $G$ around $e$. Hence, the total units charged is at most $12 \mu(G)$. Then, we prove that each edge of type 1 and 2 has been charged at least 3 units and also each edge of type 4 has been charged at least 2 units. As a result, $ 12\mu(G) \geq 3a + 3b + 2c$, which completes the proof.

We proceed with the charging scheme. For each edge $e \in M$, we investigate the structure of $G$ around $e$ and show how to do the charging.

Suppose $e \in M$ is of type 1, say $e=uv$, $\deg(u) = 1$, $\deg(v) = 4$ and $\{ u, v_1, v_2,v_3 \}$ are neighbors of $v$.
In this case, charge 3 units equally to each of the edges $uv$, $vv_1$, $vv_2$ and $vv_3$. It is obvious that regardless of the type of these edges, they have been charged at least 3 units.

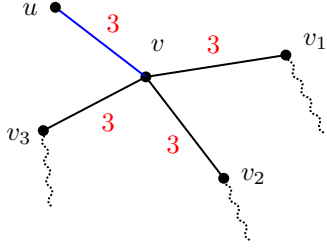
\begin{center}
    \tikzset{every picture/.style={line width=0.75pt}} 

\begin{tikzpicture}[x=0.75pt,y=0.75pt,yscale=-1,xscale=1]

\draw  [fill={rgb, 255:red, 0; green, 0; blue, 0 }  ,fill opacity=1 ] (224,124.07) .. controls (224,125.3) and (225,126.3) .. (226.23,126.3) .. controls (227.47,126.3) and (228.47,125.3) .. (228.47,124.07) .. controls (228.47,122.83) and (227.47,121.83) .. (226.23,121.83) .. controls (225,121.83) and (224,122.83) .. (224,124.07) -- cycle ;
\draw [color={rgb, 255:red, 0; green, 1; blue, 255 }  ,draw opacity=1 ]   (181.07,89.07) -- (226.23,124.07) ;
\draw    (226.23,124.07) -- (175.07,151.07) ;
\draw  [fill={rgb, 255:red, 0; green, 0; blue, 0 }  ,fill opacity=1 ] (178.83,89.07) .. controls (178.83,90.3) and (179.83,91.3) .. (181.07,91.3) .. controls (182.3,91.3) and (183.3,90.3) .. (183.3,89.07) .. controls (183.3,87.83) and (182.3,86.83) .. (181.07,86.83) .. controls (179.83,86.83) and (178.83,87.83) .. (178.83,89.07) -- cycle ;
\draw  [fill={rgb, 255:red, 0; green, 0; blue, 0 }  ,fill opacity=1 ] (172.83,151.07) .. controls (172.83,152.3) and (173.83,153.3) .. (175.07,153.3) .. controls (176.3,153.3) and (177.3,152.3) .. (177.3,151.07) .. controls (177.3,149.83) and (176.3,148.83) .. (175.07,148.83) .. controls (173.83,148.83) and (172.83,149.83) .. (172.83,151.07) -- cycle ;
\draw    (226.23,124.07) -- (296.07,113.07) ;
\draw    (226.23,124.07) -- (265.07,175.07) ;
\draw  [fill={rgb, 255:red, 0; green, 0; blue, 0 }  ,fill opacity=1 ] (262.83,175.07) .. controls (262.83,176.3) and (263.83,177.3) .. (265.07,177.3) .. controls (266.3,177.3) and (267.3,176.3) .. (267.3,175.07) .. controls (267.3,173.83) and (266.3,172.83) .. (265.07,172.83) .. controls (263.83,172.83) and (262.83,173.83) .. (262.83,175.07) -- cycle ;
\draw  [fill={rgb, 255:red, 0; green, 0; blue, 0 }  ,fill opacity=1 ] (293.83,113.07) .. controls (293.83,114.3) and (294.83,115.3) .. (296.07,115.3) .. controls (297.3,115.3) and (298.3,114.3) .. (298.3,113.07) .. controls (298.3,111.83) and (297.3,110.83) .. (296.07,110.83) .. controls (294.83,110.83) and (293.83,111.83) .. (293.83,113.07) -- cycle ;
\draw  [dash pattern={on 0.75pt off 0.75pt}]  (179.07,189.07) .. controls (177.24,187.58) and (177.06,185.92) .. (178.54,184.09) .. controls (180.03,182.26) and (179.85,180.6) .. (178.02,179.12) .. controls (176.19,177.64) and (176.01,175.98) .. (177.5,174.15) .. controls (178.98,172.32) and (178.8,170.66) .. (176.97,169.18) .. controls (175.14,167.69) and (174.96,166.03) .. (176.45,164.2) .. controls (177.94,162.37) and (177.76,160.71) .. (175.93,159.23) .. controls (174.1,157.75) and (173.92,156.09) .. (175.4,154.26) -- (175.07,151.07) -- (175.07,151.07) ;
\draw  [dash pattern={on 0.75pt off 0.75pt}]  (278.07,206.07) .. controls (275.88,205.18) and (275.24,203.64) .. (276.13,201.46) .. controls (277.02,199.28) and (276.38,197.74) .. (274.2,196.84) .. controls (272.02,195.95) and (271.38,194.41) .. (272.27,192.23) .. controls (273.16,190.05) and (272.51,188.51) .. (270.33,187.62) .. controls (268.15,186.73) and (267.51,185.19) .. (268.4,183.01) .. controls (269.29,180.83) and (268.64,179.29) .. (266.46,178.4) -- (265.07,175.07) -- (265.07,175.07) ;
\draw  [dash pattern={on 0.75pt off 0.75pt}]  (315.07,152.07) .. controls (312.84,151.3) and (312.11,149.8) .. (312.88,147.57) .. controls (313.65,145.34) and (312.92,143.85) .. (310.69,143.08) .. controls (308.46,142.31) and (307.73,140.81) .. (308.5,138.58) .. controls (309.27,136.35) and (308.54,134.86) .. (306.31,134.09) .. controls (304.08,133.32) and (303.35,131.82) .. (304.12,129.59) .. controls (304.89,127.36) and (304.16,125.87) .. (301.93,125.1) .. controls (299.7,124.33) and (298.97,122.83) .. (299.74,120.6) .. controls (300.51,118.37) and (299.78,116.88) .. (297.55,116.11) -- (296.07,113.07) -- (296.07,113.07) ;

\draw (227.23,103.7) node [anchor=north west][inner sep=0.75pt]    {$v$};
\draw (163.23,85.7) node [anchor=north west][inner sep=0.75pt]    {$u$};
\draw (205,91.4) node [anchor=north west][inner sep=0.75pt]    {$\textcolor[rgb]{1,0,0}{3}$};
\draw (202.65,140.97) node [anchor=north west][inner sep=0.75pt]    {$\textcolor[rgb]{1,0,0}{3}$};
\draw (255.15,101.97) node [anchor=north west][inner sep=0.75pt]    {$\textcolor[rgb]{1,0,0}{3}$};
\draw (235,152.4) node [anchor=north west][inner sep=0.75pt]    {$\textcolor[rgb]{1,0,0}{3}$};
\draw (303.23,100.7) node [anchor=north west][inner sep=0.75pt]    {$v_{1}$};
\draw (272.23,168.7) node [anchor=north west][inner sep=0.75pt]    {$v_{2}$};
\draw (155.23,148.7) node [anchor=north west][inner sep=0.75pt]    {$v_{3}$};

\end{tikzpicture}

\end{center}
    
Suppose $e \in M$ is of type 2, say $e=uv$, $\deg(u) = 2$, $\deg(v) = 4$, $\{u, v_1, v_2,v_3\}$ are neighbors of $v$, and $\{v,u_1\}$ are neighbors of $u$. Since $G$ does not contain parallel edges (\hyperref[property-2]{property \ref*{property-2}}), $v_i$'s are distinct. Since $G$ does not contain cycle of length three (\hyperref[property-3]{property \ref*{property-3}}), $u_1$ is also distinct from $v_i$'s.
Each of $v_i$'s have degree more than 1, because if $\deg(v_i) = 1$, we can remove $uv$ from $M$ and add $vv_i$ to it. So, the number of edges of type 1 increases, which contradicts $M$ being in $\mathcal{M}$. Using \hyperref[property-4]{property \ref*{property-4}} for vertex $v$, we conclude that all of $v_i$'s have degree 4. Using \hyperref[property-1]{property \ref*{property-1}} for $uu_1$ we conclude that $\deg(u_1) = 4$. Hence, $uu_1$ and $uv$ are of type 2 and $vv_1$,$vv_2$ and $vv_3$ are of type 3.
In this case, charge 3 units to $uu_1$, 3 units to $uv$, and 2 units to each of $vv_i$s for $1 \leq i \leq 3$. Note that all of these edges have been charged by the required value regarding their types.

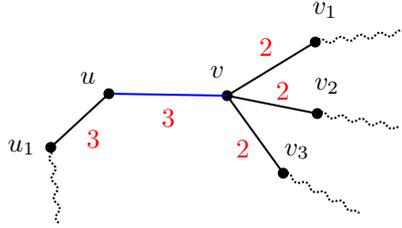
\begin{center}
    \tikzset{every picture/.style={line width=0.75pt}} 

\begin{tikzpicture}[x=0.75pt,y=0.75pt,yscale=-1,xscale=1]

\draw  [fill={rgb, 255:red, 0; green, 0; blue, 0 }  ,fill opacity=1 ] (219.83,165.07) .. controls (219.83,166.3) and (220.83,167.3) .. (222.07,167.3) .. controls (223.3,167.3) and (224.3,166.3) .. (224.3,165.07) .. controls (224.3,163.83) and (223.3,162.83) .. (222.07,162.83) .. controls (220.83,162.83) and (219.83,163.83) .. (219.83,165.07) -- cycle ;
\draw [color={rgb, 255:red, 0; green, 1; blue, 255 }  ,draw opacity=1 ]   (163.07,164.07) -- (222.07,165.07) ;
\draw    (163.07,164.07) -- (134.07,191.07) ;
\draw  [fill={rgb, 255:red, 0; green, 0; blue, 0 }  ,fill opacity=1 ] (160.83,164.07) .. controls (160.83,165.3) and (161.83,166.3) .. (163.07,166.3) .. controls (164.3,166.3) and (165.3,165.3) .. (165.3,164.07) .. controls (165.3,162.83) and (164.3,161.83) .. (163.07,161.83) .. controls (161.83,161.83) and (160.83,162.83) .. (160.83,164.07) -- cycle ;
\draw  [fill={rgb, 255:red, 0; green, 0; blue, 0 }  ,fill opacity=1 ] (131.83,191.07) .. controls (131.83,192.3) and (132.83,193.3) .. (134.07,193.3) .. controls (135.3,193.3) and (136.3,192.3) .. (136.3,191.07) .. controls (136.3,189.83) and (135.3,188.83) .. (134.07,188.83) .. controls (132.83,188.83) and (131.83,189.83) .. (131.83,191.07) -- cycle ;
\draw  [dash pattern={on 0.75pt off 0.75pt}]  (138.07,229.07) .. controls (136.24,227.58) and (136.06,225.92) .. (137.54,224.09) .. controls (139.03,222.26) and (138.85,220.6) .. (137.02,219.12) .. controls (135.19,217.64) and (135.01,215.98) .. (136.5,214.15) .. controls (137.98,212.32) and (137.8,210.66) .. (135.97,209.18) .. controls (134.14,207.69) and (133.96,206.03) .. (135.45,204.2) .. controls (136.94,202.37) and (136.76,200.71) .. (134.93,199.23) .. controls (133.1,197.75) and (132.92,196.09) .. (134.4,194.26) -- (134.07,191.07) -- (134.07,191.07) ;
\draw    (266.07,138.07) -- (222.07,165.07) ;
\draw    (267.07,174.07) -- (222.07,165.07) ;
\draw    (250.07,204.07) -- (222.07,165.07) ;
\draw  [fill={rgb, 255:red, 0; green, 0; blue, 0 }  ,fill opacity=1 ] (247.83,204.07) .. controls (247.83,205.3) and (248.83,206.3) .. (250.07,206.3) .. controls (251.3,206.3) and (252.3,205.3) .. (252.3,204.07) .. controls (252.3,202.83) and (251.3,201.83) .. (250.07,201.83) .. controls (248.83,201.83) and (247.83,202.83) .. (247.83,204.07) -- cycle ;
\draw  [fill={rgb, 255:red, 0; green, 0; blue, 0 }  ,fill opacity=1 ] (264.83,174.07) .. controls (264.83,175.3) and (265.83,176.3) .. (267.07,176.3) .. controls (268.3,176.3) and (269.3,175.3) .. (269.3,174.07) .. controls (269.3,172.83) and (268.3,171.83) .. (267.07,171.83) .. controls (265.83,171.83) and (264.83,172.83) .. (264.83,174.07) -- cycle ;
\draw  [fill={rgb, 255:red, 0; green, 0; blue, 0 }  ,fill opacity=1 ] (263.83,138.07) .. controls (263.83,139.3) and (264.83,140.3) .. (266.07,140.3) .. controls (267.3,140.3) and (268.3,139.3) .. (268.3,138.07) .. controls (268.3,136.83) and (267.3,135.83) .. (266.07,135.83) .. controls (264.83,135.83) and (263.83,136.83) .. (263.83,138.07) -- cycle ;
\draw  [dash pattern={on 0.75pt off 0.75pt}]  (266.07,138.07) .. controls (267.55,136.23) and (269.21,136.05) .. (271.04,137.53) .. controls (272.88,139.01) and (274.54,138.83) .. (276.01,136.99) .. controls (277.48,135.15) and (279.14,134.97) .. (280.98,136.45) .. controls (282.82,137.93) and (284.48,137.75) .. (285.95,135.91) .. controls (287.42,134.07) and (289.08,133.89) .. (290.92,135.37) .. controls (292.76,136.84) and (294.42,136.66) .. (295.89,134.82) .. controls (297.36,132.98) and (299.02,132.8) .. (300.86,134.28) .. controls (302.7,135.76) and (304.36,135.58) .. (305.83,133.74) .. controls (307.3,131.9) and (308.96,131.72) .. (310.8,133.2) -- (312.07,133.07) -- (312.07,133.07) ;
\draw  [dash pattern={on 0.75pt off 0.75pt}]  (267.07,174.07) .. controls (268.99,172.7) and (270.63,172.98) .. (272,174.91) .. controls (273.36,176.83) and (275,177.11) .. (276.92,175.75) .. controls (278.84,174.39) and (280.48,174.67) .. (281.85,176.59) .. controls (283.22,178.51) and (284.86,178.79) .. (286.78,177.43) .. controls (288.7,176.07) and (290.34,176.35) .. (291.71,178.27) .. controls (293.07,180.2) and (294.71,180.48) .. (296.64,179.12) .. controls (298.56,177.76) and (300.2,178.04) .. (301.57,179.96) .. controls (302.94,181.88) and (304.58,182.16) .. (306.5,180.8) -- (308.07,181.07) -- (308.07,181.07) ;
\draw  [dash pattern={on 0.75pt off 0.75pt}]  (250.07,204.07) .. controls (252.32,203.37) and (253.79,204.15) .. (254.49,206.4) .. controls (255.2,208.65) and (256.67,209.42) .. (258.92,208.72) .. controls (261.17,208.02) and (262.65,208.8) .. (263.34,211.05) .. controls (264.04,213.3) and (265.52,214.08) .. (267.77,213.38) .. controls (270.02,212.68) and (271.5,213.46) .. (272.19,215.71) .. controls (272.88,217.96) and (274.36,218.74) .. (276.61,218.04) .. controls (278.86,217.34) and (280.34,218.12) .. (281.04,220.37) .. controls (281.73,222.62) and (283.21,223.4) .. (285.46,222.7) -- (288.07,224.07) -- (288.07,224.07) ;

\draw (212.23,148.7) node [anchor=north west][inner sep=0.75pt]    {$v$};
\draw (147.23,151.7) node [anchor=north west][inner sep=0.75pt]    {$u$};
\draw (150.57,180.97) node [anchor=north west][inner sep=0.75pt]    {$\textcolor[rgb]{1,0,0}{3}$};
\draw (236.65,134.97) node [anchor=north west][inner sep=0.75pt]    {$\textcolor[rgb]{1,0,0}{2}$};
\draw (188,170.4) node [anchor=north west][inner sep=0.75pt]    {$\textcolor[rgb]{1,0,0}{3}$};
\draw (111.23,185.7) node [anchor=north west][inner sep=0.75pt]    {$u_{1}$};
\draw (245.07,156.97) node [anchor=north west][inner sep=0.75pt]    {$\textcolor[rgb]{1,0,0}{2}$};
\draw (225.07,185.97) node [anchor=north west][inner sep=0.75pt]    {$\textcolor[rgb]{1,0,0}{2}$};
\draw (263.23,116.7) node [anchor=north west][inner sep=0.75pt]    {$v_{1}$};
\draw (264.23,154.7) node [anchor=north west][inner sep=0.75pt]    {$v_{2}$};
\draw (249.23,187.7) node [anchor=north west][inner sep=0.75pt]    {$v_{3}$};

\end{tikzpicture}

\end{center}

Finally, suppose $e \in M$ is of type 4, say $e=uv$ such that $\deg(u) = \deg(v) = 4$, $\{u, v_1, v_2, v_3\}$ are neighbors of $v$, and $\{v, u_1, u_2, u_3\}$ are neighbors of $u$. Since $G$ does not contain parallel edges (\hyperref[property-2]{property \ref*{property-2}}), $v_i$'s are distinct and $u_i$'s are distinct as well. Since $G$ does not contain cycle of length three (\hyperref[property-3]{property \ref*{property-3}}), all of $u_i$'s are distinct from all of $v_i$'s.
Each of $u_i$'s have degree more than 1, because if $\deg(u_i) = 1$, we can remove $uv$ from $M$ and add $uu_i$ to it. So, the number of edges of type 1 increases which is in contradiction with $M$ being in $\mathcal{M}$. The same reasoning holds for each of $v_i$'s. As a result, all of seven edges incident to $u$ and $v$ are of type 2 and 4.

If there exists $1 \leq i,j \leq 3$ such that $u_i$ and $v_j$ are not incident to any edge of $M$, then $u_iuvv_j$ would be an augmenting path for $M$. In other words, we can remove $uv$ from $M$ and add $uu_i$ and $vv_j$ to it, obtaining a matching of size greater than $|M|$ which is in contradiction with $M$ being maximum. Hence, either all of $u_i$'s or all of $v_i$'s are incident to edges in $M$. Let's say $u_i$'s have this property.

For each $1 \leq i \leq 3$, if $\deg(u_i) = 2$ and $ u_iw \in M$ is the incident edge to $u_i$ other than $u_iu$, then $u_iw$ is of type 2 and from charging scheme of this types of edges we know that $u_iw$ have already charged 3 units to $u_iu$. Hence, there is no need to charge $u_iu$ anymore by $uv$. If $\deg(u_i) = 4$, then charge 1 unit to $uu_i$. As a result, total value charged to $uu_i$s is at most $3$ units.
For the side of $v$, for each $1 \leq i \leq 3$, if $\deg(v_i) = 2$ ($vv_i$ is of type 2), charge 3 units to $vv_i$, and if $\deg(v_i) = 4$ ($vv_i$ is of type 4), charge 2 units to $vv_i$. Using  \hyperref[property-4]{property \ref*{property-4}} for $v$, there is at most one $1 \leq i \leq 3$ such that $\deg(v_i) = 2$. 
Hence, the total value charged to $vv_i$s is at most $2+2+3 = 7$ units.
Finally, charge $2$ units to $uv$ (which is of type 4) itself. Therefore, the total charged value is at most $12$.

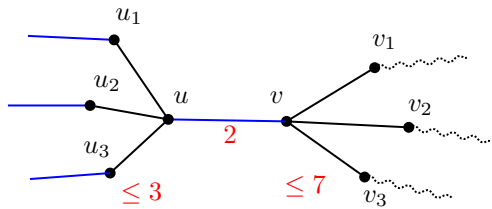
\begin{center}
    \tikzset{every picture/.style={line width=0.75pt}} 

\begin{tikzpicture}[x=0.75pt,y=0.75pt,yscale=-1,xscale=1]

\draw  [fill={rgb, 255:red, 0; green, 0; blue, 0 }  ,fill opacity=1 ] (219.83,165.07) .. controls (219.83,166.3) and (220.83,167.3) .. (222.07,167.3) .. controls (223.3,167.3) and (224.3,166.3) .. (224.3,165.07) .. controls (224.3,163.83) and (223.3,162.83) .. (222.07,162.83) .. controls (220.83,162.83) and (219.83,163.83) .. (219.83,165.07) -- cycle ;
\draw [color={rgb, 255:red, 0; green, 1; blue, 255 }  ,draw opacity=1 ]   (163.07,164.07) -- (222.07,165.07) ;
\draw    (163.07,164.07) -- (134.07,191.07) ;
\draw  [fill={rgb, 255:red, 0; green, 0; blue, 0 }  ,fill opacity=1 ] (160.83,164.07) .. controls (160.83,165.3) and (161.83,166.3) .. (163.07,166.3) .. controls (164.3,166.3) and (165.3,165.3) .. (165.3,164.07) .. controls (165.3,162.83) and (164.3,161.83) .. (163.07,161.83) .. controls (161.83,161.83) and (160.83,162.83) .. (160.83,164.07) -- cycle ;
\draw  [fill={rgb, 255:red, 0; green, 0; blue, 0 }  ,fill opacity=1 ] (131.83,191.07) .. controls (131.83,192.3) and (132.83,193.3) .. (134.07,193.3) .. controls (135.3,193.3) and (136.3,192.3) .. (136.3,191.07) .. controls (136.3,189.83) and (135.3,188.83) .. (134.07,188.83) .. controls (132.83,188.83) and (131.83,189.83) .. (131.83,191.07) -- cycle ;
\draw    (266.07,138.07) -- (222.07,165.07) ;
\draw    (283.07,168.07) -- (222.07,165.07) ;
\draw    (261.07,193.07) -- (222.07,165.07) ;
\draw  [fill={rgb, 255:red, 0; green, 0; blue, 0 }  ,fill opacity=1 ] (258.83,193.07) .. controls (258.83,194.3) and (259.83,195.3) .. (261.07,195.3) .. controls (262.3,195.3) and (263.3,194.3) .. (263.3,193.07) .. controls (263.3,191.83) and (262.3,190.83) .. (261.07,190.83) .. controls (259.83,190.83) and (258.83,191.83) .. (258.83,193.07) -- cycle ;
\draw  [fill={rgb, 255:red, 0; green, 0; blue, 0 }  ,fill opacity=1 ] (280.83,168.07) .. controls (280.83,169.3) and (281.83,170.3) .. (283.07,170.3) .. controls (284.3,170.3) and (285.3,169.3) .. (285.3,168.07) .. controls (285.3,166.83) and (284.3,165.83) .. (283.07,165.83) .. controls (281.83,165.83) and (280.83,166.83) .. (280.83,168.07) -- cycle ;
\draw  [fill={rgb, 255:red, 0; green, 0; blue, 0 }  ,fill opacity=1 ] (263.83,138.07) .. controls (263.83,139.3) and (264.83,140.3) .. (266.07,140.3) .. controls (267.3,140.3) and (268.3,139.3) .. (268.3,138.07) .. controls (268.3,136.83) and (267.3,135.83) .. (266.07,135.83) .. controls (264.83,135.83) and (263.83,136.83) .. (263.83,138.07) -- cycle ;
\draw  [dash pattern={on 0.75pt off 0.75pt}]  (266.07,138.07) .. controls (267.55,136.23) and (269.21,136.05) .. (271.04,137.53) .. controls (272.88,139.01) and (274.54,138.83) .. (276.01,136.99) .. controls (277.48,135.15) and (279.14,134.97) .. (280.98,136.45) .. controls (282.82,137.93) and (284.48,137.75) .. (285.95,135.91) .. controls (287.42,134.07) and (289.08,133.89) .. (290.92,135.37) .. controls (292.76,136.84) and (294.42,136.66) .. (295.89,134.82) .. controls (297.36,132.98) and (299.02,132.8) .. (300.86,134.28) .. controls (302.7,135.76) and (304.36,135.58) .. (305.83,133.74) .. controls (307.3,131.9) and (308.96,131.72) .. (310.8,133.2) -- (312.07,133.07) -- (312.07,133.07) ;
\draw  [dash pattern={on 0.75pt off 0.75pt}]  (283.07,168.07) .. controls (284.99,166.7) and (286.63,166.98) .. (288,168.91) .. controls (289.36,170.83) and (291,171.11) .. (292.92,169.75) .. controls (294.84,168.39) and (296.48,168.67) .. (297.85,170.59) .. controls (299.22,172.51) and (300.86,172.79) .. (302.78,171.43) .. controls (304.7,170.07) and (306.34,170.35) .. (307.71,172.27) .. controls (309.07,174.2) and (310.71,174.48) .. (312.64,173.12) .. controls (314.56,171.76) and (316.2,172.04) .. (317.57,173.96) .. controls (318.94,175.88) and (320.58,176.16) .. (322.5,174.8) -- (324.07,175.07) -- (324.07,175.07) ;
\draw  [dash pattern={on 0.75pt off 0.75pt}]  (261.07,193.07) .. controls (263.09,191.86) and (264.71,192.26) .. (265.92,194.28) .. controls (267.13,196.3) and (268.75,196.7) .. (270.77,195.49) .. controls (272.79,194.28) and (274.41,194.68) .. (275.62,196.7) .. controls (276.83,198.72) and (278.45,199.13) .. (280.47,197.92) .. controls (282.49,196.71) and (284.11,197.11) .. (285.32,199.13) .. controls (286.53,201.15) and (288.15,201.55) .. (290.17,200.34) .. controls (292.19,199.13) and (293.81,199.54) .. (295.02,201.56) .. controls (296.23,203.58) and (297.85,203.98) .. (299.87,202.77) .. controls (301.89,201.56) and (303.51,201.96) .. (304.72,203.98) -- (305.07,204.07) -- (305.07,204.07) ;
\draw [color={rgb, 255:red, 0; green, 1; blue, 255 }  ,draw opacity=1 ]   (94.07,194.07) -- (134.07,191.07) ;
\draw    (163.07,164.07) -- (124.07,157.07) ;
\draw    (163.07,164.07) -- (136.07,124.07) ;
\draw [color={rgb, 255:red, 0; green, 1; blue, 255 }  ,draw opacity=1 ]   (83.07,157.07) -- (124.07,157.07) ;
\draw [color={rgb, 255:red, 0; green, 1; blue, 255 }  ,draw opacity=1 ]   (93.07,122.07) -- (136.07,124.07) ;
\draw  [fill={rgb, 255:red, 0; green, 0; blue, 0 }  ,fill opacity=1 ] (121.83,157.07) .. controls (121.83,158.3) and (122.83,159.3) .. (124.07,159.3) .. controls (125.3,159.3) and (126.3,158.3) .. (126.3,157.07) .. controls (126.3,155.83) and (125.3,154.83) .. (124.07,154.83) .. controls (122.83,154.83) and (121.83,155.83) .. (121.83,157.07) -- cycle ;
\draw  [fill={rgb, 255:red, 0; green, 0; blue, 0 }  ,fill opacity=1 ] (133.83,124.07) .. controls (133.83,125.3) and (134.83,126.3) .. (136.07,126.3) .. controls (137.3,126.3) and (138.3,125.3) .. (138.3,124.07) .. controls (138.3,122.83) and (137.3,121.83) .. (136.07,121.83) .. controls (134.83,121.83) and (133.83,122.83) .. (133.83,124.07) -- cycle ;

\draw (212.23,148.7) node [anchor=north west][inner sep=0.75pt]    {$v$};
\draw (164.23,148.7) node [anchor=north west][inner sep=0.75pt]    {$u$};
\draw (137.83,194.47) node [anchor=north west][inner sep=0.75pt]    {$\textcolor[rgb]{1,0,0}{\leq 3}$};
\draw (119.23,173.7) node [anchor=north west][inner sep=0.75pt]    {$u_{3}$};
\draw (263.23,119.7) node [anchor=north west][inner sep=0.75pt]    {$v_{1}$};
\draw (281.23,152.7) node [anchor=north west][inner sep=0.75pt]    {$v_{2}$};
\draw (258.83,198.47) node [anchor=north west][inner sep=0.75pt]    {$v_{3}$};
\draw (124.23,140.7) node [anchor=north west][inner sep=0.75pt]    {$u_{2}$};
\draw (135.23,106.7) node [anchor=north west][inner sep=0.75pt]    {$u_{1}$};
\draw (189.07,165.97) node [anchor=north west][inner sep=0.75pt]    {$\textcolor[rgb]{1,0,0}{2}$};
\draw (219.57,190.97) node [anchor=north west][inner sep=0.75pt]    {$\textcolor[rgb]{1,0,0}{\leq 7}$};

\end{tikzpicture}

\end{center}

Note that edges in the $v$ side have been charged by their required value regarding their types. The edge $uv$, which is of type 4, has been charged 2 units. For each $1 \leq i \leq 3$, if $\deg(u_i) = 2$, then as discussed before, we know that $uu_i$ has been charged 3 units. If $\deg(u_i) = 4$, since $u_i$ is incident to an edge $e' \in M$, then $uu_i$ has been charged at least 1 unit by $e'$ and 1 unit by $uv$. Thus, all of edges in $u$ side have been charged by their required values regarding their types as well.
    
For the final argument of the proof, note that either every element of $ E(G)$ is in $M$ itself or is incident to at least one edge in $M$. Otherwise, we can add it to $M$ which is in contradiction with $M$ being maximum. As discussed in each case, edges incident to some $ e \in M $ and $e$ itself has been charged with the required value regarding their types. Hence, the proof of this lemma is done.

    \section{\tsp}\label{sec:alg2}

In this section, we present our algorithm for the $(1,2)$-TSP, detailed in \hyperref[alg4]{Algorithm 2}, and analyze its approximation factor. We also provide an explanation of how to implement this algorithm in the semi-streaming model.

\begin{algorithm}
		\caption{Our algorithm for \tsp.}
		\begin{algorithmic}[1]\label{alg4}
			\State Let $G_1$ be the subgraph of $G$ consisting of edges with weight $1$.
			\State Run \hyperref[alg1]{Algorithm 1} on $G_1$ to get a path cover $\tilde{P}$.
			\State Arbitrarily extend $\tilde{P}$ to a Hamiltonian cycle $\tilde{C}$ by adding edges between end points of $\tilde{P}$ or/and existing vertices not in $\tilde{P}$.
			\State \Return $\tilde{C}$.
		\end{algorithmic}	
	\end{algorithm}

	\begin{mytheorem}{}
		The approximation factor of \hyperref[alg4]{Algorithm 2} for $(1,2)$-TSP is $ \frac{4}{3} + \epsilon + \frac{1}{n} $.
	\end{mytheorem}
	
    \begin{myproof}
Let $ T^* $ be the optimal solution of (1,2)-TSP, $\rho^*$ be the size of an MPC in $G_1$, and $n$ be the number of vertices of $G$.
Since every Hamiltonian cycle contains $n$ edges with weights $1$ or $2$, we have $ n \leq T^* \leq 2n $. We also have
$T^*=2n - \rho^* - 1$ or $T^*= 2n-\rho^*$, where $T^*=2n - \rho^* - 1 $ occurs only when $G$ contains a Hamiltonian cycle consisting solely of edges with weight 1.

Let $\tilde{\rho}$ be the size of the path cover obtained by \hyperref[alg1]{Algorithm 1} in $G_1$.
The Hamiltonian cycle obtained by \hyperref[alg4]{Algorithm 2} has a cost of at most
$2(n-\tilde{\rho}) + \tilde{\rho} = 2n -\tilde{\rho}$.
Let $\alpha \leq 1$ be the approximation factor of \hyperref[alg1]{Algorithm 1}, then we have $ \alpha \rho^* \leq \tilde{\rho} \leq \rho^* $.
As a result, $ T^* \leq 2n-\rho^* \leq 2n - \tilde{\rho} $. We also have:
		\begin{eqnarray}\label{eq:T*}
			2n - \tilde{\rho} &\leq& 
			2n - \alpha \rho^* \nonumber =
			(2-2\alpha)n + \alpha (2n - \rho^* - 1) + \alpha \nonumber  \\
			&\leq& (2-2\alpha) T^* + \alpha T^* + \alpha \nonumber  = (2-\alpha) T^* + \alpha \nonumber \\
        &\leq& (2-\alpha) T^* + 1 \nonumber 
        \\
        &\leq& (2-\alpha) T^* + \dfrac{T^*}{n} 
        = \left(2-\alpha + \dfrac{1}{n}\right) T^*.
		\end{eqnarray}
  By \hyperref[ratio-2/3]{Theorem \ref*{ratio-2/3}}, we have $\alpha \geq \frac{2}{3} - \epsilon$. Using \hyperref[eq:T*]{Equation (\ref*{eq:T*})}, we conclude that:
\begin{eqnarray}
2n - \tilde{\rho} &\leq& 
\left(2-\alpha + \dfrac{1}{n}\right) T^* \nonumber \\ 
&\leq& \left(2-\left( \dfrac{2}{3} - \epsilon\right) + \dfrac{1}{n} \right) T^* \nonumber 
= \left(\dfrac{4}{3} + \epsilon + \dfrac{1}{n} \right) T^*. \nonumber
\end{eqnarray}
Hence, $ T^* \leq 2n - \tilde{\rho} \leq \left(\frac{4}{3} + \epsilon + \frac{1}{n} \right) T^* $. So, the approximation factor of our algorithm for (1,2)-TSP is $ \frac{4}{3} + \epsilon + \frac{1}{n} $.
\end{myproof}

\subsection{Implementation of \hyperref[alg4]{Algorithm 2} in the Semi-Streaming Model}

For a given instance of \tsp in the streaming model, we compute an approximate MPC for the induced subgraph on the edges of weight $1$ as explained in \hyperref[main1]{Theorem \ref*{main1}}, then we add extra edges to connect these paths and vertices not in these paths arbitrarily to construct a Hamiltonian cycle, which gives us a $(4/3+\epsilon+1/n)$-approximate tour for \tsp.
So, we have the main result of this section as follows.

\begin{mytheorem}{}
\label{main2}
    Given an instance of \tsp on $n$ vertices, there is a deterministic algorithm that returns a $(\frac{4}{3}+\epsilon+\frac{1}{n})$-approximate \tsp in the semi-streaming model in $O(\text{poly}(\frac{1}{\epsilon}))$ passes.   
    \end{mytheorem}

    \section{Max-TSP}\label{max-tsp}

In this section, we introduce our algorithm for Max-TSP, which closely resembles our approach for MPC. The key difference is that, instead of using $ \text{MM}_{\epsilon} $, we employ a subroutine to compute an approximate maximum weight matching in a weighted graph.

Let $\MWM$ be a subroutine for computing a $(1-\epsilon)$-approximate maximum weighted matching in a weighted graph $G$.
First, we compute a matching $M_1$ for $G$ using $\MWM$. Then, we contract the edges of $M_1$ to obtain another graph $G' = G/M_1$ and compute another matching, $M_2$, for $G'$ using $\MWM$ again. We derive the union of the two weighted matchings, $M_1 \cup M_2$. Similar to \hyperref[correct-alg1]{Lemma \ref*{correct-alg1}}, it is evident that $M_1 \cup M_2$ forms a union of vertex-disjoint paths in $G$. Finally, since the graph is complete, there can be only one vertex that is not in $M_1 \cup M_2$. In this case  we connect this vertex to one of the paths in $M_1 \cup M_2$.
Now, we add edges arbitrarily between the endpoints of the paths in $M_1 \cup M_2$ to obtain a Hamiltonian cycle $C$ for $G$. 

\begin{algorithm}
    \caption{Our algorithm for Max-TSP on a complete weighted graph $ G $.}
\begin{algorithmic}[1]\label{alg3}
    \State Run $ \text{MWM}_{\epsilon} $ on $ G $ to find a matching  $M_1$.
    \State Contract $ G $ on $ M_1 $ to get a new graph $ G^{\prime} = G/M_1 $.
    \State Run $ \text{MWM}_{\epsilon} $ on $ G^{\prime} $ to find another matching $ M_2 $.
    \State Arbitrarily extend $ M_1 \cup M_2 $ to a Hamiltonian cycle $C$ by adding edges between end points of $ M_1 \cup M_2 $
or/and existing vertices not in $ M_1 \cup M_2 $.
    \State \Return $ C $.
\end{algorithmic}	
\end{algorithm}

Note that after contracting $G$ on $M_1$ to obtain $G' = G / M_1$, this new graph might have parallel edges between to vertices. Since we aim to find a maximum matching in $G'$, we can simply consider the edge with the largest weight for parallel edges and ignore the rest.

\subsection{Analysis of the Approximation Factor of  \hyperref[alg3]{Algorithm 3}}

To analyze the approximation factor of \hyperref[alg3]{Algorithm 3}, we begin with a series of lemmas.

\begin{mylemma}{}\label{M-in-C}
    Suppose $ C $ is a cycle of length $k$ in a weighted graph $ G $. Then, there exists a matching $ M \subseteq C $ such that $w(M) \geq \frac{k-1}{2k}w(C) $.
\end{mylemma}

\begin{myproof}
Assume that $e \in C$ is the edge with the minimum weight. Hence, $w(e) \leq w(C)/k$.
Since $C - e$ is a path, there is a matching $M \subseteq C - e$ (which is also a subset of $C$) whose weight is at least $w(C-e)/2$. Finally,
$$ w(M) \geq \frac{1}{2}w(C-e) = \frac{1}{2}(w(C)-w(e)) \geq \frac{1}{2} \left(w(C)-\frac{w(C)}{k} \right) = \frac{k-1}{2k}w(C).$$
\end{myproof}

\begin{mylemma}{}\label{M-geq-1/3}
 Suppose $T$ is a path or a cycle in a weighted graph $G$. Then there exists a matching $M \subseteq T $ such that $w(M) \geq \frac{1}{3}w(T)$.
\end{mylemma}

\begin{myproof}
We have two cases
\begin{itemize}
    \item $T$ is a path. Enumerate the edges of $T$ from one end point to the other. The odd numbered edges form a matching called $M_{\text{odd}}$. The same applies for even numbered edges, which form a matching called $M_{\text{even}}$. Since $T = M_{\text{odd}} \cup M_{\text{even}}$, at least one of these two matchings has weight no less than $w(T)/2$.
    \item $T$ is a cycle. If it is a cycle of length $2$ (i.e. $T$ consists of two parallel edges), then obviously we can pick the edge $e$ with bigger weight that satisfies $w(e) \geq \frac{1}{2}w(T) \geq \frac{1}{3}w(T)$.
    If the length of $T$ is at least $3$, then using \hyperref[M-in-C]{Lemma \ref*{M-in-C}}, we conclude that there is a matching $M \subseteq T$ such that
    $w(M) \geq \frac{k-1}{2k}w(T) \geq \frac{1}{3}w(T)$.
\end{itemize}
\end{myproof}

Now, we provide a lemma similar to
\hyperref[mu>(rho-M)/3]{Lemma \ref*{mu>(rho-M)/3}} which works for the weighted version.

\begin{mylemma}{}\label{approx-1/2+(1-s)/4}
 Suppose $M$ is a matching in a weighted graph $G$ and $ C^*$ is a maximum weight Hamiltonian cycle of $ G $. Then,
$$ \mu(G/M) \geq \frac{w(C^*) - w(M)}{6} - \frac{1}{3n}w(C^*) . $$
\end{mylemma}

\begin{myproof}
 We contract $ C^* $ on $ M $ in two steps. First, we contract $ C^* $ on the edges in  $C^* \cap M$. Next, we contract the resulting graph on $M^\prime = M \setminus C^*$.
After the first step, $C^\prime = C^* / (C^* \cap M)$ is a cycle with weight $w(C^*) - w(C^* \cap M)$ (see \hyperref[fig-approx-1/2+1/2s-10]{Figure \ref*{fig-approx-1/2+1/2s-10}}). 

 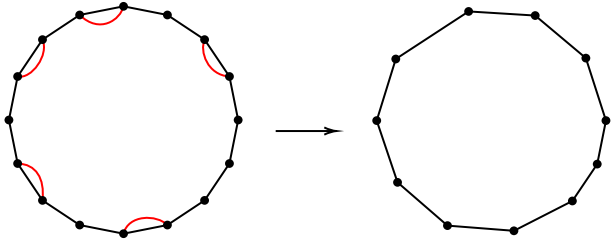
\begin{figure}[h]
			    
		\begin{center}
			\tikzset{every picture/.style={line width=0.75pt}} 
			
			\begin{tikzpicture}[x=0.75pt,y=0.75pt,yscale=-0.55,xscale=0.55]

\draw    (197.24,38.91) -- (230.97,61.45) ;
\draw    (253.51,95.19) -- (261.42,134.98) ;
\draw [color={rgb, 255:red, 0; green, 0; blue, 0 }  ,draw opacity=1 ]   (261.42,134.98) -- (253.51,174.77) ;
\draw    (197.24,231.04) -- (157.45,238.96) ;
\draw [color={rgb, 255:red, 0; green, 0; blue, 0 }  ,draw opacity=1 ]   (83.92,208.5) -- (61.38,174.77) ;
\draw    (53.47,134.98) -- (61.38,95.19) ;
\draw    (83.92,61.45) -- (117.65,38.91) ;
\draw [color={rgb, 255:red, 0; green, 0; blue, 0 }  ,draw opacity=1 ]   (117.65,38.91) -- (157.45,31) ;
\draw [color={rgb, 255:red, 0; green, 0; blue, 0 }  ,draw opacity=1 ]   (61.38,95.19) -- (83.92,61.45) ;
\draw [color={rgb, 255:red, 0; green, 0; blue, 0 }  ,draw opacity=1 ]   (230.97,61.45) -- (253.51,95.19) ;
\draw [color={rgb, 255:red, 0; green, 0; blue, 0 }  ,draw opacity=1 ]   (197.24,231.04) -- (230.97,208.5) ;
\draw [color={rgb, 255:red, 0; green, 0; blue, 0 }  ,draw opacity=1 ]   (117.65,231.04) -- (157.45,238.96) ;
\draw [color={rgb, 255:red, 255; green, 0; blue, 0 }  ,draw opacity=1 ]   (117.65,38.91) .. controls (136.04,56.84) and (154.57,43.14) .. (157.45,31) ;
\draw [color={rgb, 255:red, 255; green, 0; blue, 0 }  ,draw opacity=1 ]   (61.38,95.19) .. controls (73.99,97.94) and (90.11,75.37) .. (83.92,61.45) ;
\draw [color={rgb, 255:red, 255; green, 0; blue, 0 }  ,draw opacity=1 ]   (61.38,174.77) .. controls (78.02,175.3) and (86.08,188.19) .. (83.92,208.5) ;
\draw [color={rgb, 255:red, 255; green, 0; blue, 0 }  ,draw opacity=1 ]   (230.97,61.45) .. controls (226.29,72.15) and (233.54,93.91) .. (253.51,95.19) ;
\draw [color={rgb, 255:red, 255; green, 0; blue, 0 }  ,draw opacity=1 ]   (157.45,238.96) .. controls (161.82,221.23) and (187.61,221.23) .. (197.24,231.04) ;
\draw    (530.88,39.44) -- (576.85,78.37) ;
\draw    (576.85,78.37) -- (595.07,135.51) ;
draw [color={rgb, 255:red, 0; green, 0; blue, 0 }  ,draw opacity=1 ]   (595.07,135.51) -- (587.15,175.3) ;
\draw    (387.11,135.51) -- (404.41,79.18) ;
\draw    (404.41,79.18) -- (470.49,35.67) ;
\draw [color={rgb, 255:red, 0; green, 0; blue, 0 }  ,draw opacity=1 ]   (511.58,236.32) -- (564.61,209.03) ;
\draw [color={rgb, 255:red, 0; green, 0; blue, 0 }  ,draw opacity=1 ]   (451.3,231.57) -- (511.58,236.32) ;
\draw    (295.47,145.07) -- (348.47,145.07) ;
\draw [shift={(350.47,145.07)}, rotate = 180] [color={rgb, 255:red, 0; green, 0; blue, 0 }  ][line width=0.75]    (10.93,-3.29) .. controls (6.95,-1.4) and (3.31,-0.3) .. (0,0) .. controls (3.31,0.3) and (6.95,1.4) .. (10.93,3.29)   ;

\draw    (564.61,209.03) -- (587.15,175.3) ;

\draw    (451.3,231.57) -- (406, 192);

\draw    (470.49,35.67) -- (530.88,39.44);

\draw    (595.07,135.51) -- (587.15,175.3) ;

\fill   (406, 192) circle [x radius= 4, y radius= 4];
\fill   (595.07,135.51) circle [x radius= 4, y radius= 4];
\fill   (587.15,175.3) circle [x radius= 4, y radius= 4];
\fill   (530.88,39.44) circle [x radius= 4, y radius= 4];
\fill   (576.85,78.37) circle [x radius= 4, y radius= 4];
\fill   (470.49,35.67) circle [x radius= 4, y radius= 4];
\fill   (404.41,79.18) circle [x radius= 4, y radius= 4];
\fill   (451.3,231.57) circle [x radius= 4, y radius= 4];
\fill   (387.11,135.51) circle [x radius= 4, y radius= 4];
\fill   (511.58,236.32) circle [x radius= 4, y radius= 4];
\fill   (564.61,209.03) circle [x radius= 4, y radius= 4];

\draw    (387.11,135.51) -- (406, 192) ;

\draw    (197.24,38.91) -- (157.45,31);
\draw    (253.51,174.77) -- (230.97,208.5);
\draw    (117.65,231.04) -- (83.92,208.5);

\fill   (197.24,38.91) circle [x radius= 4, y radius= 4];
\fill   (230.97,61.45) circle [x radius= 4, y radius= 4];
\fill   (253.51,95.19) circle [x radius= 4, y radius= 4];
\fill   (261.42,134.98) circle [x radius= 4, y radius= 4];
\fill   (230.97,208.5) circle [x radius= 4, y radius= 4];
\fill   (253.51,174.77) circle [x radius= 4, y radius= 4];
\fill   (197.24,231.04) circle [x radius= 4, y radius= 4];
\fill   (157.45,238.96) circle [x radius= 4, y radius= 4];
\fill   (83.92,208.5) circle [x radius= 4, y radius= 4];
\fill   (117.65,231.04) circle [x radius= 4, y radius= 4];
\fill    (61.38,174.77)  circle [x radius= 4, y radius= 4];
\fill  (53.47,134.98)  circle [x radius= 4, y radius= 4];
\fill   (83.92,61.45) circle [x radius= 4, y radius= 4];
\fill   (61.38,95.19) circle [x radius= 4, y radius= 4];
\fill   (117.65,38.91) circle [x radius= 4, y radius= 4];
\fill   (157.45,31) circle [x radius= 4, y radius= 4];

\draw    (53.47,134.98) -- (61.38,174.77) ;

			\end{tikzpicture}
		\end{center}
  
   \caption{$ C^* $ remains a path cover after contraction on $\textcolor[rgb]{1,0,0}{C^* \cap M}$ (red edges).}\label{fig-approx-1/2+1/2s-10}
						\end{figure}
    
    Here $ M^\prime $ is a matching that connects some vertices of $ C^\prime $ together (see \hyperref[fig-approx-1/2+(1-s)/4-1]{Figure \ref*{fig-approx-1/2+(1-s)/4-1}}, \hyperref[fig-approx-1/2+(1-s)/4-2]{Figure \ref*{fig-approx-1/2+(1-s)/4-2}} and \hyperref[fig-approx-1/2+(1-s)/4-3]{Figure \ref*{fig-approx-1/2+(1-s)/4-3}}).\\
    Assume that the length of $C'$ is $k$. 
    Using \hyperref[M-in-C]{Lemma \ref*{M-in-C}}, there is a matching  $ M^* \subseteq C' $ whose weight is at least $\frac{k-1}{2k} w(C')$
    (see \hyperref[fig-approx-1/2+(1-s)/4-4]{Figure \ref*{fig-approx-1/2+(1-s)/4-4}}).
    Since the matching $M_1$ contains at most half of the edges of $C^*$, we conclude that $k \geq n/2$. As a result, 
 \begin{eqnarray}\label{eq:m>(1-s)/2p}
     w(M^*) &\geq& \frac{k-1}{2k} w(C') \geq 
    \frac{n-2}{2n} (w(C^*) - w(C^* \cap M) ) \nonumber \\
    &\geq& \frac{w(C^*) - w(C^* \cap M)}{2} - \frac{1}{n} w(C^*).
 \end{eqnarray}
	Since $ M^\prime \cap C^\prime = \emptyset $, we conclude that $ M^\prime \cap M^* = \emptyset$. Because $ M^* $ and $ M^\prime $ are matchings, it follows that $ M^* \cup M^\prime $ is a union of disjoint paths and cycles (see \hyperref[fig-approx-1/2+(1-s)/4-5]{Figure \ref*{fig-approx-1/2+(1-s)/4-5}}).
    As a result, after doing the second step of contraction, $ M^* / M^\prime $ would also be a disjoint union of paths and cycles (whose number of edges is equal to $|M^*|$ since $M^* \cap M^\prime = \emptyset$) in $ C^\prime / M^\prime $. For instance, if $ M^* \cup M^\prime $ contains a cycle of length $4$, then $ M^* / M^\prime $ would contain a cycle of length $2$ which contains parallel edges.\\
    Now, consider each connected component of $ M^* / M^\prime $. This component is either a path or a cycle. Hence, by \hyperref[M-geq-1/3]{Lemma \ref*{M-geq-1/3}}, We obtain a matching with a weight of at least one-third of the weight of the component.

\input{figures/fig4}

Finally, since these components are vertex-disjoint, the union of obtained matching would be a matching whose weight is at least $w(M^*)/3$.
Note that this matching is also a matching in $G/M$. Hence, using \hyperref[eq:m>(1-s)/2p]{Equation \ref*{eq:m>(1-s)/2p}}, we have
\begin{eqnarray*}
  \mu(G/M) &\geq& \frac{w(M^*)}{3} \geq \frac{w(C^*) - w(C^* \cap M)}{6} - \frac{1}{3n}w(C^*) \\
    &\geq& \frac{w(C^*) - w(M)}{6} - \frac{1}{3n}w(C^*).
\end{eqnarray*}
\centering

\end{myproof}

So, we have the following theorem which is a lower bound for the approximation factor of 
\hyperref[alg3]{Algorithm 3}.

\begin{mytheorem}{}\label{ratio-7/12}
	The approximation factor of \hyperref[alg3]{Algorithm 3} is at least $ 
  \left( \frac{7}{12} - \frac{3}{4n} \right) (1 - \epsilon) $. 
\end{mytheorem}

\begin{myproof}
Let $C^*$ be a maximum weight Hamiltonian cycle in $G$. By \hyperref[M-in-C]{Lemma \ref*{M-in-C}}, there exists at least one matching $M \subseteq C^* $ whose weight is at least 
$$ \frac{n-1}{2n} w(C^*). $$
Since $M_1$ is a $(1-\epsilon)$-approximation of the maximum weighted matching in $G$ we have 
$$ w(M_1) \geq \frac{(1-\epsilon)(n-1)}{2n} w(C^*) .$$
By using \hyperref[approx-1/2+(1-s)/4]{Lemma \ref*{approx-1/2+(1-s)/4}} for $M_2$ on $G/M_1$, we have
 \begin{eqnarray*}
        w(M_1 \cup M_2) &=& w(M_1) + w(M_2) \\
        &\geq& 
        w(M_1) + (1-\epsilon)\mu(G/M_1) \\
        &\geq& 
        w(M_1) + \frac{1-\epsilon}{6} (w(C^*) - w(M_1)) - \dfrac{1-\epsilon}{3n}w(C^*) \\
        &=&
        (1-\epsilon)\left( \dfrac{1}{6} - \dfrac{1}{3n} \right) w(C^*) + \frac{5 + \epsilon}{6} w(M_1) \\
        &\geq&
        (1-\epsilon)\left( \dfrac{1}{6} - \dfrac{1}{3n} \right) w(C^*) + \frac{5}{6} w(M_1) \\
        & \geq &
        (1-\epsilon)\left( \frac{1}{6} - \dfrac{1}{3n} \right) w(C^*) + \frac{5(1-\epsilon)(n-1)}{12n} w(C^*) \\
        &=&
        (1-\epsilon)\left( \dfrac{1}{6} - \dfrac{1}{3n}  + \dfrac{5}{12}  - \dfrac{5}{12n}  \right) w(C^*) \\
        &=&
        \left( \dfrac{7}{12} - \dfrac{3}{4n} \right) (1 - \epsilon ) w(C^*).
    \end{eqnarray*}

Since the weight of the edges of $G$ are nonnegative, we have
$$ w(C) \geq w(M_1 \cup M_2) \geq \left( \frac{7}{12} - \frac{3}{4n} \right) (1 - \epsilon ) w(C^*). $$
Finally, $C$ is a Hamiltonian cycle which means $w(C^*) \geq w(C) $. Hence, the approximation factor of \hyperref[alg1]{Algorithm 1} is at least $\left( \frac{7}{12} - \frac{3}{4n} \right) (1 - \epsilon )$. 
\end{myproof}

\subsection{Implementation of  \hyperref[alg3]{Algorithm 3} in the semi-streaming model}
The implementation of \hyperref[alg3]{Algorithm 3} in the semi-streaming model follows a similar approach as described in the previous section. Therefore, we omit a detailed explanation here. However, note that in this part, we should use a subroutine for computing a $(1-\epsilon)$-approximate maximum weight matching in the semi-streaming model. First, we recall the following theorem from \cite{huangS23}. We use the algorithm of this theorem as $\MWM$ in our semi-streaming implementation of \hyperref[alg3]{Algorithm 3}.

\begin{mytheorem}(Theorem 1.3 in \cite{huangS23})
\label{smw}
There exists a deterministic
algorithm that returns a $(1-\epsilon)$-approximate maximum weight matching using $\text{poly}(\frac{1}{\epsilon})$ passes in the semi-streaming model.
The
algorithm requires $O(n\cdot \log W \cdot \text{poly}(\frac{1}{\epsilon}))$ words of memory where $W$ is the maximum edge weight in the graph.
\end{mytheorem}

Thus, \hyperref[alg3]{Algorithm 3}, \hyperref[ratio-7/12]{Theorem \ref*{ratio-7/12}}, and \hyperref[smw]{Theorem \ref*{smw}} present the following theorem for Max-TSP in the semi-streaming model.

\begin{mytheorem}{}
\label{main3}
    Given an instance of Max-TSP on $n$ vertices, there is an algorithm that returns a $(\frac{7}{12} - \frac{3}{4n})(1-\epsilon)$-approximate Max-TSP the semi-streaming model in $O(\text{poly}(\frac{1}{\epsilon}))$ passes. The
algorithm requires $O(n\cdot \log W \cdot \text{poly}(\frac{1}{\epsilon}))$ words of memory where $W$ is the maximum edge weight in the graph.
\end{mytheorem}
   
    \section{Future Work}\label{repeat}

As a future work we propose the following algorithm that can help to improve the approximation factor for MPC in the semi-streaming model.
The algorithm improves \hyperref[alg1]{Algorithm 1} by iteratively finding new matchings and contracting the graph over these matchings.
 It is crucial to ensure that this process preserves the path cover property. Hence, during the $k$th iteration of our loop, we must remove all the edges in $G$ that are incident to a middle point of any path (connected component) within $ \cup_{i=1}^k M_i$. This is because $ \left( \cup_{i=1}^k M_i \right) \cup M_{k+1}$ must remain a path cover, which means $M_{k+1}$ cannot include any edge incident to a middle point of a path in $ \cup_{i=1}^k M_i$.
See \hyperref[alg-repeat]{Algorithm 4}.
\begin{algorithm}
\label{moreiterations}
\caption{Extension of \hyperref[alg1]{Algorithm 1}.}
\begin{algorithmic}[1]\label{alg-repeat}
    \State Run $ \text{MM}_{\epsilon} $ (or $\MWM$ for weighted version) on $ G $ to find a matching  $M_1$.
    \State Let $i = 1$.
    \State while $M_i \neq \emptyset$:
    
    \State \quad Let $G^{(i)} = G$.
    \State \quad Remove all $e \in E(G^{(i)}) \setminus (\cup_{k=0}^i M_k)$ from $E(G^{(i)})$ that are incident to at least one middle point of a path (connected component) in $\cup_{k=0}^i M_k$.
    \State \quad Contract $G^{(i)}$ on   $\cup_{k=0}^i M_k$.
    \State \quad Run $ \text{MM}_{\epsilon} $ (or $\MWM$ for weighted version) on $ G^{(i)} $ to find a matching  $M_{i+1}$.
    \State \quad $i=i+1$
    \State \Return $ \cup_{k=1}^{ i } M_k $.
\end{algorithmic}
\end{algorithm}

\begin{figure}[h]

 \hspace{0.01\linewidth}
	\subcaptionbox{$\textcolor[rgb]{1,0,0}{M_1}$ (red edges). \label{fig-upper-many-iterations-1}}%
	[.25\linewidth]{\centering
\tikzset{every picture/.style={line width=0.75pt}} 

\begin{tikzpicture}[x=0.75pt,y=0.75pt,yscale=-1,xscale=1]

\draw [color={rgb, 255:red, 255; green, 0; blue, 0 }  ,draw opacity=1 ]   (95.98,90.53) -- (154.37,71.56) ;
\draw [color={rgb, 255:red, 255; green, 0; blue, 0 }  ,draw opacity=1 ]   (95.98,151.94) -- (154.37,170.91) ;
\draw [color={rgb, 255:red, 0; green, 0; blue, 0 }  ,draw opacity=1 ]   (190.47,121.23) -- (154.37,170.91) ;
\draw [color={rgb, 255:red, 0; green, 0; blue, 0 }  ,draw opacity=1 ]   (154.37,71.56) -- (190.47,121.23) ;
\draw    (154.37,71.56) -- (154.37,170.91) ;
\draw  [fill={rgb, 255:red, 0; green, 0; blue, 0 }  ,fill opacity=1 ] (152.14,71.56) .. controls (152.14,70.32) and (153.14,69.32) .. (154.37,69.32) .. controls (155.61,69.32) and (156.61,70.32) .. (156.61,71.56) .. controls (156.61,72.79) and (155.61,73.79) .. (154.37,73.79) .. controls (153.14,73.79) and (152.14,72.79) .. (152.14,71.56) -- cycle ;
\draw  [fill={rgb, 255:red, 0; green, 0; blue, 0 }  ,fill opacity=1 ] (188.23,121.23) .. controls (188.23,120) and (189.23,119) .. (190.47,119) .. controls (191.7,119) and (192.7,120) .. (192.7,121.23) .. controls (192.7,122.47) and (191.7,123.47) .. (190.47,123.47) .. controls (189.23,123.47) and (188.23,122.47) .. (188.23,121.23) -- cycle ;
\draw  [fill={rgb, 255:red, 0; green, 0; blue, 0 }  ,fill opacity=1 ] (93.74,90.53) .. controls (93.74,89.3) and (94.74,88.3) .. (95.98,88.3) .. controls (97.21,88.3) and (98.21,89.3) .. (98.21,90.53) .. controls (98.21,91.76) and (97.21,92.76) .. (95.98,92.76) .. controls (94.74,92.76) and (93.74,91.76) .. (93.74,90.53) -- cycle ;
\draw  [fill={rgb, 255:red, 0; green, 0; blue, 0 }  ,fill opacity=1 ] (152.14,170.91) .. controls (152.14,169.68) and (153.14,168.68) .. (154.37,168.68) .. controls (155.61,168.68) and (156.61,169.68) .. (156.61,170.91) .. controls (156.61,172.14) and (155.61,173.14) .. (154.37,173.14) .. controls (153.14,173.14) and (152.14,172.14) .. (152.14,170.91) -- cycle ;
\draw  [fill={rgb, 255:red, 0; green, 0; blue, 0 }  ,fill opacity=1 ] (93.74,151.94) .. controls (93.74,150.7) and (94.74,149.7) .. (95.98,149.7) .. controls (97.21,149.7) and (98.21,150.7) .. (98.21,151.94) .. controls (98.21,153.17) and (97.21,154.17) .. (95.98,154.17) .. controls (94.74,154.17) and (93.74,153.17) .. (93.74,151.94) -- cycle ;

\end{tikzpicture}

  }
 \hspace{0.08\linewidth}
	\subcaptionbox{$\textcolor[rgb]{1,0,0}{M_2}$ (red edges). \label{fig-upper-many-iterations-2}}%
	[.25\linewidth]{\centering
\tikzset{every picture/.style={line width=0.75pt}} 

\begin{tikzpicture}[x=0.75pt,y=0.75pt,yscale=-1,xscale=1]

\draw [color={rgb, 255:red, 0; green, 0; blue, 0 }  ,draw opacity=1 ]   (190.47,121.23) -- (154.37,170.91) ;
\draw [color={rgb, 255:red, 0; green, 0; blue, 0 }  ,draw opacity=1 ]   (154.37,71.56) -- (190.47,121.23) ;
\draw [color={rgb, 255:red, 255; green, 0; blue, 0 }  ,draw opacity=1 ]   (154.37,71.56) -- (154.37,170.91) ;
\draw  [fill={rgb, 255:red, 0; green, 0; blue, 0 }  ,fill opacity=1 ] (152.14,71.56) .. controls (152.14,70.32) and (153.14,69.32) .. (154.37,69.32) .. controls (155.61,69.32) and (156.61,70.32) .. (156.61,71.56) .. controls (156.61,72.79) and (155.61,73.79) .. (154.37,73.79) .. controls (153.14,73.79) and (152.14,72.79) .. (152.14,71.56) -- cycle ;
\draw  [fill={rgb, 255:red, 0; green, 0; blue, 0 }  ,fill opacity=1 ] (188.23,121.23) .. controls (188.23,120) and (189.23,119) .. (190.47,119) .. controls (191.7,119) and (192.7,120) .. (192.7,121.23) .. controls (192.7,122.47) and (191.7,123.47) .. (190.47,123.47) .. controls (189.23,123.47) and (188.23,122.47) .. (188.23,121.23) -- cycle ;
\draw  [fill={rgb, 255:red, 0; green, 0; blue, 0 }  ,fill opacity=1 ] (152.14,170.91) .. controls (152.14,169.68) and (153.14,168.68) .. (154.37,168.68) .. controls (155.61,168.68) and (156.61,169.68) .. (156.61,170.91) .. controls (156.61,172.14) and (155.61,173.14) .. (154.37,173.14) .. controls (153.14,173.14) and (152.14,172.14) .. (152.14,170.91) -- cycle ;

\end{tikzpicture}

  }
 \hspace{0.08\linewidth}
	\subcaptionbox{MPC (blue edges). \label{fig-upper-many-iterations-3}}%
	[.25\linewidth]{\centering
\tikzset{every picture/.style={line width=0.75pt}} 

\begin{tikzpicture}[x=0.75pt,y=0.75pt,yscale=-1,xscale=1]

\draw [color={rgb, 255:red, 0; green, 38; blue, 255 }  ,draw opacity=1 ]   (95.98,90.53) -- (154.37,71.56) ;
\draw [color={rgb, 255:red, 0; green, 38; blue, 255 }  ,draw opacity=1 ]   (95.98,151.94) -- (154.37,170.91) ;
\draw [color={rgb, 255:red, 0; green, 38; blue, 255 }  ,draw opacity=1 ]   (190.47,121.23) -- (154.37,170.91) ;
\draw [color={rgb, 255:red, 0; green, 38; blue, 255 }  ,draw opacity=1 ]   (154.37,71.56) -- (190.47,121.23) ;
\draw    (154.37,71.56) -- (154.37,170.91) ;
\draw  [fill={rgb, 255:red, 0; green, 0; blue, 0 }  ,fill opacity=1 ] (152.14,71.56) .. controls (152.14,70.32) and (153.14,69.32) .. (154.37,69.32) .. controls (155.61,69.32) and (156.61,70.32) .. (156.61,71.56) .. controls (156.61,72.79) and (155.61,73.79) .. (154.37,73.79) .. controls (153.14,73.79) and (152.14,72.79) .. (152.14,71.56) -- cycle ;
\draw  [fill={rgb, 255:red, 0; green, 0; blue, 0 }  ,fill opacity=1 ] (188.23,121.23) .. controls (188.23,120) and (189.23,119) .. (190.47,119) .. controls (191.7,119) and (192.7,120) .. (192.7,121.23) .. controls (192.7,122.47) and (191.7,123.47) .. (190.47,123.47) .. controls (189.23,123.47) and (188.23,122.47) .. (188.23,121.23) -- cycle ;
\draw  [fill={rgb, 255:red, 0; green, 0; blue, 0 }  ,fill opacity=1 ] (93.74,90.53) .. controls (93.74,89.3) and (94.74,88.3) .. (95.98,88.3) .. controls (97.21,88.3) and (98.21,89.3) .. (98.21,90.53) .. controls (98.21,91.76) and (97.21,92.76) .. (95.98,92.76) .. controls (94.74,92.76) and (93.74,91.76) .. (93.74,90.53) -- cycle ;
\draw  [fill={rgb, 255:red, 0; green, 0; blue, 0 }  ,fill opacity=1 ] (152.14,170.91) .. controls (152.14,169.68) and (153.14,168.68) .. (154.37,168.68) .. controls (155.61,168.68) and (156.61,169.68) .. (156.61,170.91) .. controls (156.61,172.14) and (155.61,173.14) .. (154.37,173.14) .. controls (153.14,173.14) and (152.14,172.14) .. (152.14,170.91) -- cycle ;
\draw  [fill={rgb, 255:red, 0; green, 0; blue, 0 }  ,fill opacity=1 ] (93.74,151.94) .. controls (93.74,150.7) and (94.74,149.7) .. (95.98,149.7) .. controls (97.21,149.7) and (98.21,150.7) .. (98.21,151.94) .. controls (98.21,153.17) and (97.21,154.17) .. (95.98,154.17) .. controls (94.74,154.17) and (93.74,153.17) .. (93.74,151.94) -- cycle ;

\end{tikzpicture}

  }
\hfill
 \caption{An example of a graph where \hyperref[alg-repeat]{Algorithm 4} terminates after two iterations. The algorithm produces a $\dfrac{3}{4}$-approximation of the Maximum Path Cover (MPC).}\label{fig-last-example}
\end{figure}
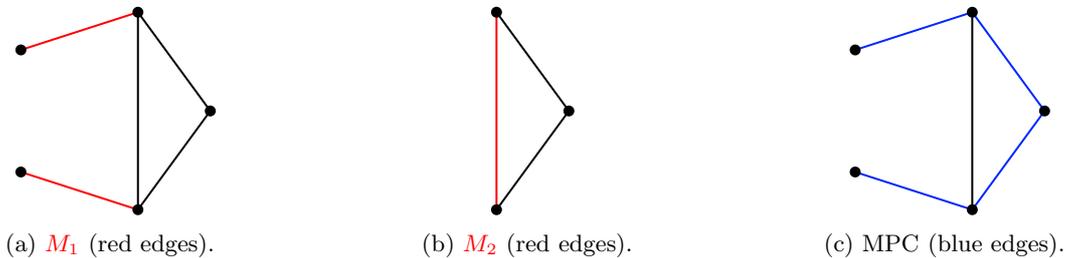

We leave the computation of the approximation factor of \hyperref[alg-repeat]{Algorithm 4} as a challenging open problem. Currently, we know that the approximation factor is at most $3/4$. Consider the graph in \hyperref[fig-upper-many-iterations-1]{Figure \ref*{fig-upper-many-iterations-1}}: the algorithm may select the red edges as $M_1$. After contraction, it might select the red edge in \hyperref[fig-upper-many-iterations-2]{Figure \ref*{fig-upper-many-iterations-2}} as $M_2$. In the next iteration, the graph becomes empty, as we must remove any edge incident to a middle point of $M_1 \cup M_2$. Thus, the algorithm terminates with a path of length 3. However, the MPC has 4 edges (see \hyperref[fig-upper-many-iterations-3]{Figure \ref*{fig-upper-many-iterations-3}}).

The main bottleneck to find the approximation ratio of \hyperref[alg-repeat]{Algorithm 4} is that after the second iteration, there might be edges incident to the middle points of $M_1 \cup M_2$ that should not be contained in the next matching. Otherwise $M_1 \cup M_2 \cup M_3$ would not be a path cover. Hence, we should remove these edges from the contracted graph in order to make sure that the union of matchings remains a valid path cover.
While running line 5 of \hyperref[alg-repeat]{Algorithm 4}, a bunch of edges might be removed.
We are not aware of any argument to bound the number of these edges.
This prevents us to provide a guarantee like \hyperref[mu>(rho-M)/3]{Lemma \ref*{mu>(rho-M)/3}} and \hyperref[approx-1/2+(1-s)/4]{Lemma \ref*{approx-1/2+(1-s)/4}}.
Finding the exact approximation ratio of \hyperref[alg-repeat]{Algorithm 4} seems to require clever new ideas, already for three matchings.

\bibstyle{plainurl}    
\bibliography{main.bib}

\end{document}